**Title:** Investigating Use of Low-Cost Sensors to Increase Accuracy and Equity of Real-Time Air Quality Information

**Authors:** Ellen M. Considine[1]*, Danielle Braun[1,2], Leila Kamareddine[1], Rachel C. Nethery[1]†, and Priyanka deSouza[3]†

**Affiliations:**
[1] Department of Biostatistics, Harvard T.H. Chan School of Public Health, Boston, Massachusetts, 02115, USA.
[2] Department of Data Science, Dana-Farber Cancer Institute, Boston, Massachusetts, 02215, USA.
[3] Department of Urban and Regional Planning, University of Colorado Denver, Denver, Colorado, 80202, USA.
† R.C.N. and P.D. contributed equally to this paper.
* Corresponding author: ellen_considine@g.harvard.edu

**Graphical Abstract:** *included as a separate pdf file so that it can be resized more easily.*

**Abstract:**
Environmental Protection Agency (EPA) air quality (AQ) monitors, the "gold standard" for measuring air pollutants, are sparsely positioned across the US. Low-cost sensors (LCS) are increasingly being used by the public to fill in the gaps in AQ monitoring; however, LCS are not as accurate as EPA monitors. In this work, we investigate factors impacting the differences between an individual's true (unobserved) exposure to air pollution and the exposure reported by their nearest AQ instrument (which could be either an LCS or an EPA monitor). We use simulations based on California data to explore different combinations of hypothetical LCS placement strategies (e.g. at schools or near major roads), for different numbers of LCS, with varying plausible amounts of LCS device measurement error. We illustrate how real-time AQ reporting could be improved (or, in some cases, worsened) by using LCS, both for the population overall and for marginalized communities specifically. This work has implications for the integration of LCS into real-time AQ reporting platforms.

**Synopsis:** How does real-time AQ information change as we deploy AQ sensors with different accuracies in different numbers and places?

## MAIN TEXT

### 1. Introduction

Decades of research have documented the adverse health impacts of both short- and long-term exposure to air pollution. In this study, we focus on fine particulate matter ($PM_{2.5}$), an air pollutant that has been associated with various adverse health outcomes [1–3]. In the US, although air quality (AQ) across the years has been improving overall, disparities between $PM_{2.5}$ concentrations experienced by subpopulations persist [4–6]. In addition, certain parts of the US are experiencing, or are likely to experience, higher air pollution (including $PM_{2.5}$ exposure) from climate-change related events and processes such as exacerbated wildfires and dust storms [7]. To develop effective AQ management plans and address key concerns of equity, accurate and high resolution AQ monitoring data are needed.

The Federal Reference Method or Federal Equivalence Method (FRM or FEM) monitors deployed by the US Environmental Protection Agency (EPA) are considered the "gold standard" for measuring AQ. However, due to the prohibitively high capital (USD $10,000+) and operating costs [8] of these instruments, they have been deployed in less than a third of all US counties [9]. Even in counties that have an EPA monitor, the often substantial within-county variability in AQ can render measurements unrepresentative of the pollution levels experienced by many residents of the county [10]. As these monitors tend to be deployed in more populous locations [11,12], residents in rural areas tend to be farther away from monitors.

In recent years, low-cost AQ sensors (< USD $2,500 as defined by the EPA Air Sensor Guidebook [13], but often much cheaper) have been gaining attention as a supplement to FRM/FEM monitoring, and many are deployed by private citizens in addition to public and private entities [14]. Key motivations of individuals (ascertained from online product reviews) include managing health impacts of wildfire smoke and other air pollution as well as detecting air pollution sources of concern [15]. Networks of these sensors can help increase the spatial resolution and frequency of AQ measurements [16]. However, the measurements from low-cost sensors (henceforth, LCS) have lower accuracy than EPA monitors, and can be further affected by environmental conditions such as relative humidity, temperature, and aerosol composition [14,17–19]. Recent work has shown that algorithmic correction can reduce the error in LCS, but does not eliminate the problem [20–23]. These studies have also highlighted the challenge of developing corrections that are transferable across measurements collected in different locations and/or time periods. Despite these drawbacks, research has shown that LCS can be useful in identifying air pollution hotspots, expanding local awareness about AQ and health, and alerting the public of short-term changes in AQ, which may facilitate reduction of exposure to air pollution, e.g. by individuals choosing not to exercise outdoors on a high pollution day [13,24–26].

To our knowledge, the question of how incorporating measurements from LCS into real-time reporting systems affects the accuracy of people's AQ information remains unanswered. In addition to sensor measurement error, an important factor that must be considered is the spatial distribution (both spatial density and relative placement) of LCS. In this study, we assume that individuals view AQ data from the nearest instrument – EPA monitor or LCS – (e.g., as shown on a smartphone app) as their current AQ exposure. Despite the potential for LCS to increase and democratize access to AQ information, a 2021 study found that locations of PurpleAir (one of the most widely used LCS brands, costing around $250 per sensor) in the US tend to be disproportionately located in neighborhoods that have higher incomes and higher percentages of white residents, compared both to the locations of EPA monitors and to the US overall [14]. This suggests that, relative to more privileged groups, residents of marginalized neighborhoods (who, as previously noted, tend to experience worse air quality) may have less access to information about their local AQ, a precursor to adaptive actions which could be taken to protect their health.

In summary, while collection and dissemination of LCS data have been increasing, there is a need for evaluating the impact of integrating LCS data into AQ reporting platforms. In this study, we investigate how altering the number of LCS deployed, amount and type of sensor measurement error, and relative placements of LCS affects the accuracy of daily AQ information available to individuals from their nearest AQ instrument (defined to be either an LCS or an EPA monitor). Our main objectives are to (a) increase the nuance with which various groups (scientists, policy makers, community organizations, etc.) can think about and discuss tradeoffs when it comes to measuring and reporting AQ to the public and (b) to suggest directions for future work to both the LCS instrument / data science and environmental health / policy communities. We focus on $PM_{2.5}$ and base our LCS investigation on PurpleAir sensors, the data from which is used in a number of popular regional real-time AQ maps [27,28].

Our analysis consists first of simulating realistic LCS $PM_{2.5}$ measurements under numerous hypothetical LCS deployment and sensor measurement error scenarios in the state of California. Then, we compare the local AQ information available to individuals in each simulation scenario with that produced by (i) EPA monitors only, as well as (ii) the existing PurpleAir sensor network. We dedicate special attention to evaluating how each scenario impacts disparities in AQ information accuracy for marginalized groups, such as those living in communities with high rates of poverty or with high proportions of nonwhite or Hispanic residents. Our findings can be used to inform decisions about (a) where to place LCS to make real-time AQ reporting more accurate and equitable, (b) how many LCS to deploy, (c) whether existing sensor calibration approaches yield sufficient accuracy to justify use of LCS for real-time AQ reporting, and (d) what amount of error is "tolerable" for future LCS deployments.

In section 2 (Materials & Methods), we describe each step of the analysis and the sources of data used. Section 3 (Results) includes a comparison of the impact of different types and amounts of sensor measurement error for LCS at current PurpleAir locations, as well as a comparison across different LCS placement strategies and numbers of LCS deployed. Section 4 (Discussion) includes conclusions, limitations, and some ideas for future investigation.

## 2. Materials & Methods

2.1 Study Setting and Overview

To evaluate the potential impact of LCS measurements on localized AQ information accuracy, our study leverages real data on EPA monitor locations, PurpleAir LCS locations, and sociodemographic and geospatial features in California. Our choice to situate the study in California was primarily motivated by California's widespread LCS uptake (California contains over half of the US's PurpleAir sensors), in part prompted by concerns about increasing air pollution from wildfires [29–32].

To ensure that our simulations accounted for realistic spatial and temporal variability in $PM_{2.5}$, we assumed the "true" (error-free and comprehensive) daily ambient $PM_{2.5}$ concentrations were those obtained from an ensemble model predicting $PM_{2.5}$ exposures daily at 1 km x 1 km in 2016, created by Di et al. (2016 was the most recent year for which these predictions were available) [33]. These estimates agreed well with ground-based reference measurements: the 10-fold cross-validated $R^2$ was 0.86 for the US overall and 0.80 for the Pacific coast states (including California). We also considered the current locations of EPA monitors (n=154 in California) to be fixed, and in our simulations, we set the $PM_{2.5}$ measurements from each of these monitors to be equal to the Di et al. $PM_{2.5}$ estimates at these locations (i.e., we assumed that there was no error in the measurements from these monitors).

Details of our simulation procedure are provided in the following subsections. Here is a brief overview to serve as a roadmap:

1. Using the placement strategy specified by the simulation scenario, select hypothetical locations of LCS
2. For each 1 km x 1 km grid in California, identify the grid centroid's nearest AQ instrument (among the real EPA monitor locations and hypothetical LCS locations)
3. For each day in 2016, simulate LCS $PM_{2.5}$ measurements by adding simulated device measurement error to the "true" $PM_{2.5}$ estimates from Di et al. at each hypothetical LCS's location
4. Evaluate the accuracy and equity of AQ information observed (based on the measurements from the nearest AQ instrument) across all 475,772 grids in California and 366 days in 2016

2.2 Selecting LCS Locations

Table S1 (in the Supporting Information) describes the data sets, data processing steps, and sampling methods used to select locations for LCS in each simulation.

To guide hypothetical LCS deployment strategies focused on environmentally and socially marginalized communities, we leveraged the CalEnviroScreen (CES) index, developed by the California Office of Environmental Health Hazard Assessment, which describes both environmental and socioeconomic-demographic marginalization at the Census tract level [34], as well as its environmental component, henceforth referred to as the Pollution Score. The Pollution Score incorporates data on air pollution (ozone, $PM_{2.5}$, diesel PM emissions, toxic chemical releases from facilities) and traffic density; pesticides, groundwater threats, impaired water bodies, and drinking water contamination index; solid waste, hazardous waste, and cleanup sites. The Pearson correlation between the Pollution Scores and the $PM_{2.5}$ estimates from Di et al. is 0.48. The socioeconomic-demographic disadvantage index used by CES incorporates data on asthma, low birth weight, cardiovascular disease, education, linguistic isolation, poverty, unemployment, and housing burden. The CES Score is a product of these environmental and socioeconomic-demographic indices.

In this study, we considered the following five hypothetical LCS placement strategies, illustrated in Figure S1: (a) at randomly selected real outdoor PurpleAir locations, (b) at randomly selected public schools, (c) at randomly selected locations favoring proximity to major roads, (d) at randomly selected locations favoring high CES Score, and (e) at randomly selected locations favoring high Pollution Score. For the last three placement strategies, "favoring" refers to weighted random sampling (respectively using nearby road lengths, CES Score, and Pollution Score as weights) to determine placement locations. We compared each of these placement strategies across different numbers of sensors deployed (0, 50, 100, 250, 500, and 1000 LCS to show the trends). To provide context, average numbers of LCS assigned to the Los Angeles, Sacramento, Imperial counties (a large city, a medium-small city, and a well-known environmental justice focus area) under each placement strategy are provided in Table S2.

2.3 Observing AQ Information from the Nearest AQ Instrument

We assumed that all individuals in each 1 km x 1 km grid in California observed daily AQ measurements from the instrument (either EPA monitor or LCS based on simulated placement strategy) nearest to their grid centroid, as shown in Figure 1. In the True Air Pollution

Exposure column (left), the background color in each grid cell represents the "true" air pollution that individuals experience (obtained from the Di et al. estimates). Note that these true exposures are most often not observed. The colors inside the triangles and circles represent the measurements from EPA monitors and LCS, respectively. LCS measurement error is represented in the bottom row, where the color inside the circle differs from the background color of the grid it's in. In the Shown Air Pollution Exposure column (right), the background color in each grid is the air pollution measurement that individuals observe from their nearest AQ instrument. The differences between the AQ that individuals experience and the AQ that they are shown are indicated by the red and blue X's in Figure 1. The red X's indicate cells where AQ is over-classified, i.e., the AQ shown to residents is worse than the AQ truly experienced. The blue X's indicate cells where AQ is under-classified, i.e., the AQ shown to residents is better than the AQ truly experienced.

Figure 1 illustrates three distinct sources of AQ reporting error to which we will refer throughout the rest of this paper: (i) distance to the nearest AQ instrument, (ii) local variability in air quality, and (iii) sensor measurement error. As an example of distance-based errors, the distance between an individual in A5 and the nearest AQ instrument is large, so an individual in A5 is unlikely to be shown accurate measurements of their air pollution exposure (in Figure 1, they would be shown that their exposure is 15 $\mu$g/m3 instead of the true exposure, which is 5 $\mu$g/m3). As an example of local variability-based errors, while an individual in C5 is close to an EPA monitor (D5), local variability in AQ between C5 and D5 results in misclassification of C5's AQ (they would be shown that their exposure is 15 $\mu$g/m3, while their true exposure is 5 $\mu$g/m3). Even if a cell contains an LCS, sensor measurement error may still result in reporting error, as in the case of an individual in C2: under the setting of device measurement error, they would be shown that their exposure is 30 $\mu$g/m3, instead of their true exposure, which is 50 $\mu$g/m3. These effects can also co-occur, as for an individual in D2: the nearest AQ instrument is in cell C2, which, in addition to having lower air pollution than D2, also suffers from LCS measurement error. In this study, we help disentangle these effects.

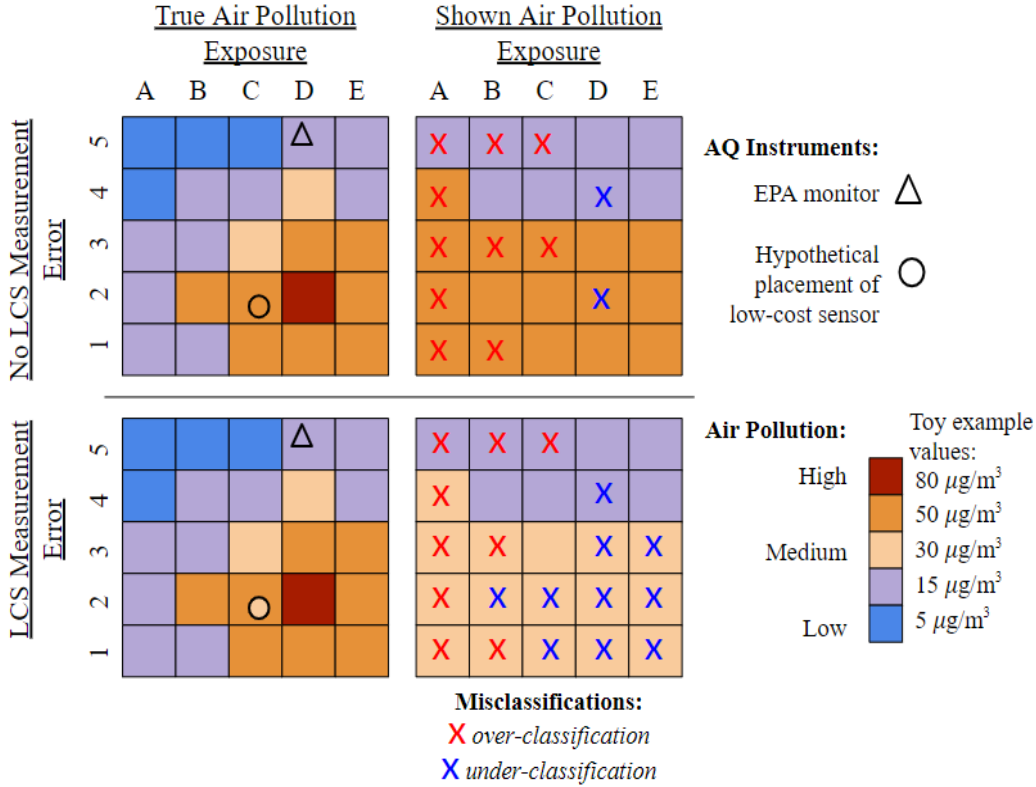

Figure 1. AQ Information Reporting Diagram. An illustration of the assumed process of AQ information reporting (from each grid's nearest AQ instrument), under two scenarios: without LCS measurement error (top row of panels) and with LCS measurement error (bottom row of panels). In the True Exposure column, the background color of each grid represents the (unobserved) air quality an individual in that grid experiences, whereas in the Shown Exposure column, the background color represents the air quality an individual in that grid observes from their nearest AQ instrument.

2.4 Simulating Sensor Measurement Error

In each hypothetical (simulated) LCS deployment scenario, daily $PM_{2.5}$ "measurements" from LCS were generated by adding sensor measurement error to the Di et al. $PM_{2.5}$ estimates, in several different ways. First, we selected measurement error distributions informed by a tiered target for AQ instrument accuracy proposed at an EPA workshop [35]. The proposal is that AQ measurements (i) for regulatory purposes require accuracy of ±10% of the true average $PM_{2.5}$ in that area, (ii) for mapping spatial gradients and monitoring microenvironments require accuracy of ±25%, and (iii) for hotspot detection require accuracy of ±50%. In our simulations, sensor measurement errors were generated both (a) differentially with respect to true $PM_{2.5}$ and (b) non-differentially (with error magnitude not varying across $PM_{2.5}$ levels). The former was motivated by empirical LCS observations, and accounts for the possibility of some spatial and temporal correlation in the sensor measurement errors due to some spatial and temporal smoothing induced by the Di et al. modeling approach. The latter assumes independence of the sensor measurement errors (post calibration of the LCS).

Second, we simulated LCS measurement errors in a manner enabling assessment of a nation-wide correction algorithm for PurpleAir sensors developed by EPA researchers [36]. Specifically, we sampled errors from the empirical distribution of residuals obtained by comparing measurements from EPA monitors in California to the corrected measurements from collocated PurpleAir sensors.

These procedures are detailed in the Supplemental Notes in the SI. Comparison of the characteristics and effects of all these different types and amounts of sensor measurement error is facilitated by Table 1 in the Results.

2.5 Evaluating the AQ Information

The final step of each simulation was to evaluate the error between the true AQ exposures (which are most often unobserved) and the exposures reported by the nearest AQ instrument, summarized across all the grids and days. We evaluated accuracy in AQ reporting using the mean absolute error (MAE) in reported $PM_{2.5}$ concentrations and the misclassification rates of the U.S. AQ index, or AQI [37].

The AQI classifies AQ into six levels, with different public health recommendations for each. Green = "Good", Yellow = "Moderate", Orange = "Unhealthy for Sensitive Groups", Red = "Unhealthy", Purple = "Very Unhealthy", and Maroon = "Hazardous". AQI is often reported as a combination of air pollutants, however, for this analysis we used the single-pollutant version for $PM_{2.5}$ [38]. We hypothesize that most people use these classifications to inform their activity rather than the exact concentrations of $PM_{2.5}$ (or any other air pollutant), so we calculated the percent of over- and under-classifications, the percent of misclassifications greater than one level (e.g. Orange → Green or Yellow → Red), and what we term Unhealthy-Healthy Misclassifications (UHM): the fraction of days that a healthy (H) classification is shown, out of the days that are truly unhealthy (U). This last metric may be of the most concern for public health. For this dichotomous variable, we defined Green and Yellow to be healthy, and Orange through Maroon to be unhealthy.

When calculating these metrics, weighting each grid by its population density allows us to evaluate the accuracy of AQ information available to individuals in California. We also performed the calculations unweighted by population density, which represents averaging across the land area instead of averaging across individuals. However, as the population-weighted metrics are more relevant for public policy (e.g., for health impact assessments), we focus on these results in the main text; the unweighted results are provided in the SI.

For each combination of sensor placement strategy, number of sensors, and type/amount of sensor measurement error, we ran 100 replicates and averaged the metrics across them to account for random sampling variability. The results are robust to the number of replicates (i.e., 50 vs 100).

2.6 Equity Analysis

In addition to the overall population metrics (averaging across all grids in California), we calculated the metrics for marginalized subsets of the population, to determine if certain sensor placement strategies resulted in more equitable or less equitable access to AQ information.

We obtained socio-demographic features from the 2016 American Community Survey [39] using the R package tidycensus [40], at the finest spatial resolution for which they were available: Census block group (CBG) level for race/ethnicity and population density, and Census tract level for socioeconomic status. To merge these features with the AQ data, we performed an overlay of the CBG shapefile with the 1 km x 1 km grid centroids from the Di et al. estimates. Any block group or tract that did not contain a grid centroid was ignored. Although this procedure tends to exclude more CBGs with smaller land area and higher population density, our main analysis weighting by population density counteracts possible bias.

For this analysis, we used percentage of non-white individuals (all but non-Hispanic white) for the percent marginalized by race/ethnicity and the percent of people living under the poverty line for those marginalized by socioeconomic status. Our decision to use one minus the percent of non-Hispanic white people to represent disadvantage by race/ethnicity (elsewhere referred to as "% nonwhite" for verbal simplicity) was informed by a preliminary calculation showing that Hispanic white people on average experience higher pollution and socioeconomic disadvantage than the overall white population. We defined CBGs with high % nonwhite and high % poverty to be CBGs that fell into the top quintiles of these measures across the 1 km x 1 km grid centroids (≥58.1% and ≥23.5% respectively). For the 0.3% of CBGs that were missing data on % nonwhite, we substituted in Census tract-level % nonwhite. Data for % poverty was only available at the tract level. Only for one tract with population > 0 (tract 6037920200 with population 5,000) were we unable to retrieve tract-level data, and thus we omitted that tract from the analysis.

In summary, for the equity analysis we calculated the AQ reporting error in CBGs with high % poverty and high % nonwhite residents. Maps of % poverty and % nonwhite residents in California are provided in Figure S2.

2.7 Software Availability

Code used to download and process the datasets, run the analyses, and generate the figures and tables can be found at https://github.com/EllenConsidine/LCS_placement_sims

## 3. Results

3.1 Descriptive Statistics

Summary statistics of average annual $PM_{2.5}$ (according to the 1km x 1km estimates by Di et al.), % poverty, CES Scores, % nonwhite, and population density, for population subgroups as well as locations (centroids of 1 km x 1 km grid cells) targeted in each of the LCS placement strategies, are shown in Table S4. One key observation, consistent with the environmental justice literature [4–6], is that CBGs with high % poverty or high % nonwhite have higher annual average $PM_{2.5}$ than the population overall. As shown in Figure S3, these marginalized subgroups also experience far more days classified as unhealthy by the AQI (level orange and higher) than the overall population. One differentiator between the two marginalized subgroups (high % nonwhite and high % poverty) is that CBGs with high % nonwhite tend to have higher population density than CBGs with high % poverty.

Another observation from Table S4, consistent with external findings [14], is that among real EPA monitor locations, PurpleAir locations, and the hypothetical LCS placement strategies considered, PurpleAir locations have by far the lowest % poverty. By contrast, EPA monitor locations have higher % poverty than any of the LCS placement strategies considered. Among the LCS placement strategies considered, schools and locations favored by CES Score have the highest % poverty. LCS placements at schools also have the highest % nonwhite out of any of the EPA monitor, PurpleAir, or hypothetical LCS locations.

S8

Finally, while schools and PurpleAir locations tend to be in CBGs with higher population density, locations chosen to favor proximity to roads, CES Score, and Pollution Score tend to have lower population density.

### 3.2 Comparing Different Types and Amounts of Sensor Measurement Error, Assuming LCS are Placed at Current PurpleAir Locations

The average distance to the nearest EPA monitor is 10.11 km for the population overall, 8.94 km for CBGs with high % nonwhite residents, and 8.69 km for CBGs with high % poverty. When we include LCS at all current locations of outdoor PurpleAir sensors, the average distance to the nearest AQ instrument drops dramatically, to 2.41 km, 2.49 km, and 2.82 km, respectively. Note that these results, like all those in the main text, are weighted by population density.

Table 1 summarizes how the accuracy of daily AQ information changes when we compare the scenario where people only have access to EPA monitors, with the scenario where people have access to EPA monitors and LCS at current locations of outdoor PurpleAir sensors. Under the scenario with LCS at PurpleAir locations, we compare the different sensor measurement error types, as described in the Methods. The first column of Table 1 shows the amount of sensor measurement error under each measurement error type (calculated as the standard deviation of the mean-zero simulated errors). For all the differential sensor measurement error scenarios (10%, 25%, and empirical residual-based), marginalized subgroups on average experience higher sensor measurement error because their air pollution exposure is higher.

Next, we use several metrics to describe the accuracy of observed AQ information under each scenario: (a) absolute error (deviation from the true exposure value), captured using the mean absolute error (MAE) and 95th percentile (to illustrate the upper end of the error distribution in addition to the mean), (b) rate of misclassification (either over- or under-classification) of the AQI, and (c) rate of Unhealthy-Healthy misclassifications (UHM), which we define as the fraction of days with unhealthy AQI that are misreported as healthy AQI.

Table 1 shows that when the LCS have no sensor measurement error (i.e. they are as accurate as EPA monitors in our simulations), deploying them at all the real locations of PurpleAir sensors roughly halves the MAE and 95th percentile of error in daily reported air quality. These improvements are smaller for CBGs with high % poverty, likely because PurpleAir sensors tend to be situated in more socioeconomically privileged areas.

Counterintuitively, even in the absence of sensor measurement error, placement of LCS at current PurpleAir locations leads to increases in the rates of under-classification of the AQI and UHMs. These reductions in classification accuracy are likely due to local variability in AQ and the fact that PurpleAir locations have lower annual average $PM_{2.5}$ than the state overall (Table S4). This issue is exacerbated by sensor measurement error.

With non-differential sensor measurement error, LCS (at all the current PurpleAir locations) with error magnitudes of ±10% and ±25% both generally improve on the no-LCS case except for CBGs with high % poverty, where the MAE increases slightly. The impact of sensor measurement error is likely exacerbated for these CBGs with high % poverty due to the socioeconomic bias of PurpleAir locations. By contrast, while 10% differential sensor measurement error improves on the no-LCS case in terms of absolute error, 25% differential error and empirical residual-based error worsen the real-time AQ reporting for all groups and by all metrics, except for some small reductions in over-classification of the AQI.

Table 1. Comparing Impacts of Different Sensor Accuracies. Results (weighted by population density) when there are no LCS vs. LCS at all real PurpleAir locations (n = 4,343), assuming different kinds and amounts of sensor measurement error (ME), averaged across 100 simulation replicates to account for randomness in the sensor measurement error generation. Unless otherwise specified, "errors" refer to the difference between the true exposure experienced at each grid centroid and the exposure reported from the nearest AQ instrument. AQI under-classification is when the true exposure class is greater than what someone is shown, and over-classification is when the true exposure class is less than what someone is shown. Rate of UH misclassification (UHM) is the fraction of days with unhealthy AQI (Orange+) that are misreported as healthy AQI (Green or Yellow).

| Type/Amount of Sensor Measurement Error | Std. Dev. of Sensor Measurement Error ($\mu g/m^3$) | MAE ($\mu g/m^3$) | 95th Percentile of Errors ($\mu g/m^3$) | Under-classified AQI (%) | Over-classified AQI (%) | UHM (%) |
|---|---|---|---|---|---|---|
| Overall Population | | | | | | |
| No LCS (only EPA monitors) | — | 1.46 | 4.45 | 2.05 | 6.79 | 11.37 |
| No Sensor Error | 0 | 0.79 | 2.79 | 2.12 | 2.37 | 15.02 |
| 10% Non-differential | 0.5 | 0.94 | 2.89 | 2.41 | 2.80 | 15.51 |
| 25% Non-differential | 1.25 | 1.33 | 3.52 | 3.07 | 4.13 | 16.27 |
| 10% Differential | 0.88 | 1.10 | 3.29 | 3.19 | 3.61 | 20.11 |
| 25% Differential | 2.19 | 1.85 | 5.39 | 5.04 | 6.45 | 28.38 |
| EPA Correction Residual Decile Draws | 3.32 | 2.45 | 7.16 | 8.27 | 6.02 | 27.10 |
| Population living in CBGs with high % nonwhite | | | | | | |
| No LCS (only EPA monitors) | — | 1.34 | 4.14 | 2.15 | 6.53 | 10.35 |
| No Sensor Error | 0 | 0.75 | 2.59 | 2.22 | 2.48 | 13.98 |
| 10% Non-differential | 0.5 | 0.89 | 2.71 | 2.53 | 2.95 | 14.57 |
| 25% Non-differential | 1.25 | 1.29 | 3.39 | 3.31 | 4.38 | 15.51 |
| 10% Differential | 0.92 | 1.08 | 3.21 | 3.45 | 3.84 | 19.87 |
| 25% Differential | 2.30 | 1.90 | 5.53 | 5.66 | 6.88 | 28.52 |
| EPA Correction Residual Decile Draws | 3.46 | 2.51 | 7.40 | 9.43 | 6.33 | 27.28 |

| Population living in CBGs with high % poverty | | | | | | |
|---|---|---|---|---|---|---|
| No LCS (only EPA monitors) | — | 1.28 | 4.06 | 2.09 | 5.79 | 8.40 |
| No Sensor Error | 0 | 0.81 | 2.80 | 2.22 | 2.68 | 10.06 |
| 10% Non-differential | 0.5 | 0.94 | 2.90 | 2.53 | 3.10 | 10.36 |
| 25% Non-differential | 1.25 | 1.31 | 3.52 | 3.31 | 4.40 | 11.38 |
| 10% Differential | 0.97 | 1.13 | 3.41 | 3.46 | 3.91 | 15.99 |
| 25% Differential | 2.44 | 1.94 | 5.75 | 5.78 | 6.71 | 25.55 |
| EPA Correction Residual Decile Draws | 3.65 | 2.54 | 7.59 | 9.60 | 6.16 | 24.20 |

These results highlight the potential for (i) LCS to reduce both the distance to the nearest AQ instrument and the absolute error in daily $PM_{2.5}$ reporting, (ii) the accuracy of classification-based AQ information to diverge from the accuracy of concentration-based AQ information, (iii) different AQ information outcomes for different subsets of the population, which is related to LCS placement characteristics, and (iv) the dependence of these insights on the type and amount of sensor measurement error.

3.3 Additionally Comparing Placement Strategies and Numbers of Sensors Deployed

We now discuss how different hypothetical LCS placement strategies and numbers of LCS deployed affect access to real-time AQ information under both non-differential and differential sensor measurement error scenarios.

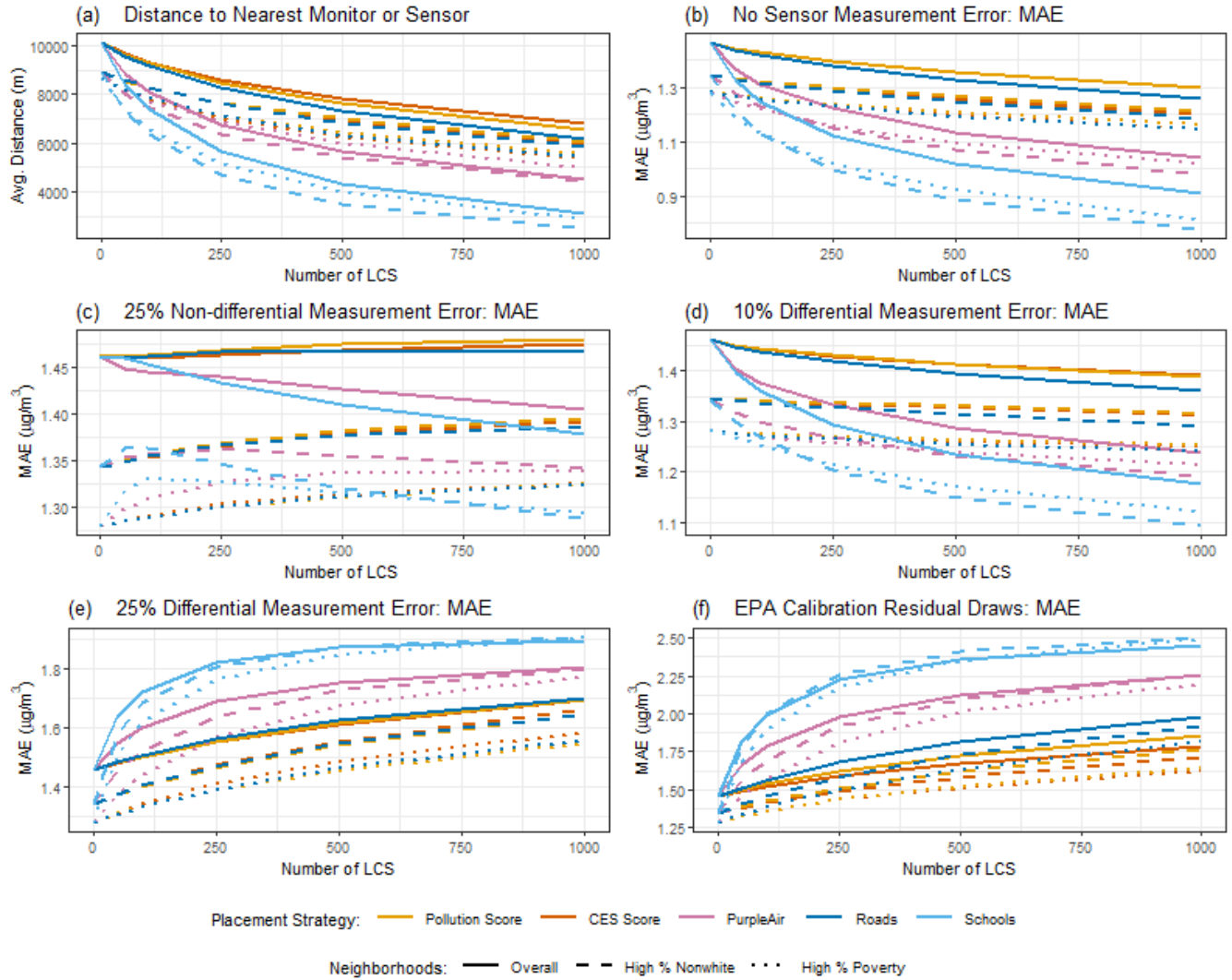

Figure 2. Distance and MAE. Distance to the nearest AQ instrument and mean absolute error (between what is reported vs. experienced) resulting from different numbers of LCS deployed, LCS placement strategies, and sensor measurement error types and amounts. All results were calculated using 366 days and averaged across 100 simulation replicates, weighted by population density. Panel a shows the distance to the nearest AQ instrument (monitor or LCS), panel b shows MAE when there is no LCS device measurement error, panel c shows MAE when LCS device measurement error is 25% non-differential, panel d shows MAE when LCS device measurement error is 10% differential, panel e shows MAE when LCS device measurement error is 25% differential, and panel f shows MAE when LCS device measurement error is sampled from the empirical distribution of residuals from PurpleAir LCS measurements (calibrated with the EPA equation) compared with collocated EPA monitor measurements.

Figure 2 shows the average distance to the nearest AQ instrument as well as the MAE resulting from simulations under each hypothetical LCS deployment scenario (i.e., each combination of placement strategy and number of LCS deployed) and for each sensor measurement error type. The vertical scales of each plot are different, to facilitate close inspection of the lines. The plot for 10% non-differential measurement error is not shown because it is very similar to panel b (the results with no sensor measurement error).

The six panels in Figure 2 help distinguish between the three different components of error in real-time AQ reporting. First, we observe that with zero or low amounts of LCS measurement error, much of the average error in daily AQ reporting is due to individuals' distance to the nearest AQ instrument, as illustrated by the similarity in line ordering and slopes between panels a, b, and d. Another observation, however, is that the impact of reduction in distance to the nearest AQ instrument can be confounded by local variability in AQ. For example, while LCS placements favoring CBGs with high Pollution Score result in greater reduction in distance to the nearest AQ instrument than placements favoring CBGs with high CES Score (panel a), the CES Score placement strategy results in lower MAE (panel b; the solid yellow line hides the solid orange line). This phenomenon is explained by local variability in AQ because the Pollution Score primarily highlights areas with large local sources of pollution, so measurements in those pollution "hotspots" may not be representative of air pollution even in nearby communities.

For low amounts of sensor measurement error (i.e. 10% non-differential and differential, as shown in panel d), reductions in daily AQ reporting error due to decreased distance to the nearest AQ instrument mitigate the impact of the LCS measurement error, improving MAEs across the board. When sensor measurement error is increased to 25% non-differential (panel c), deploying LCS only improves MAE under certain placement strategies: at schools and in current PurpleAir locations. This is largely because these placement strategies prioritize areas with higher population density.

With 25% differential and empirical residual-based sensor measurement errors (panels e and f), the impact of reduced distance to the nearest AQ instrument is overshadowed by the increased error in the sensor measurements, worsening AQ reporting across the board. Under these large amounts of sensor measurement error, placements favoring high CES Score result in the least error in AQ reporting. In nearly all cases, marginalized subgroups experience lower MAE than the population overall.

Figure 3 shows the rate of UHMs for all placement strategies, numbers of LCS deployed, and sensor measurement error types and amounts. One of the most noticeable patterns from Figure 3 is how the rate of UHMs increases when any LCS are introduced under all placement strategies and sensor measurement error types. This is especially true for LCS placements based on CES Score and Pollution Score (for the population overall and CBGs with high % nonwhite): the UHM rates stay basically constant despite the changing number of LCS. We posit that this is due to high local variability in AQ in Census tracts with high Pollution or CES Scores.

Another important observation is that although marginalized subgroups experience a higher number of unhealthy days in absolute terms, the fraction of those unhealthy days misclassified as healthy is lower than for the population overall. And, crucially, the CES and Pollution Score-based placements lead to the lowest UHM rates for CBGs with high % poverty. However, school locations are the only placement strategy resulting in decreasing UHM rates for large numbers of LCS deployed (when sensor measurement error is nondifferential, in panels a through c). A similar reversal is observed in panel c of Figure 2 for marginalized subgroups. To investigate whether the school placement strategy might produce the lowest rate of UHMs for many LCS deployed, we ran simulations with LCS at all the schools (n=7,548). The results are summarized in Table S6. Notably, with 10% nondifferential sensor measurement error, this produces lower UHM rates than the EPA monitors alone. With 25% nondifferential sensor measurement error, the UHM rates are only slightly higher (e.g., 11.8% vs 11.4% for the population overall).

One last note on Figure 3 is that although generally increasing sensor measurement error increases the rate of UHMs, the empirical residual-based errors (panel f) result in slightly lower UHM rates than the 25% differential scenario (in panel e) for LCS at schools and PurpleAir locations.

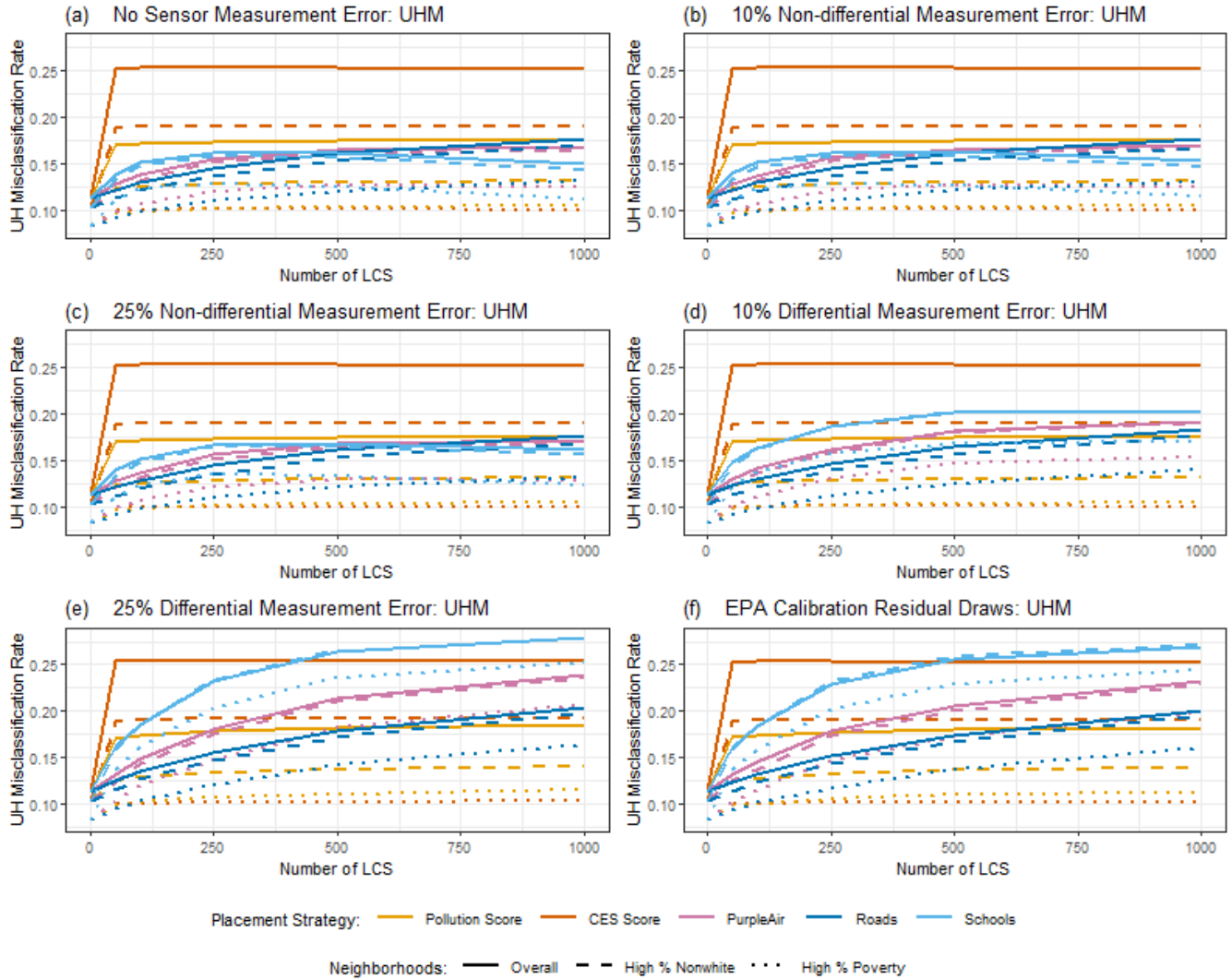

Figure 3. Unhealthy-Healthy Misclassifications. Rates of UHMs resulting from different numbers of LCS deployed, LCS placement strategies, and sensor measurement error types and amounts. UH misclassification occurs when the AQ is unhealthy (Orange+) but is reported as healthy (Green or Yellow); the UHM rate is calculated by dividing the fraction of UH misclassifications by the total fraction of unhealthy days experienced by each group. All results were calculated using 366 days and averaged across 100 simulation replicates, weighted by population density. Panel a shows UHM rate when there is no LCS device measurement error, panel b shows UHM rate when LCS device measurement error is 10% non-differential, panel c shows UHM rate when LCS device measurement error is 25% non-differential, panel d shows UHM rate when LCS device measurement error is 10% differential, panel e shows UHM rate when LCS device measurement error is 25% differential, and panel f shows UHM rate when LCS device

measurement error is sampled from the empirical distribution of residuals from PurpleAir LCS measurements (calibrated with the EPA equation) compared with collocated EPA monitor measurements.

Further insight can be gained from Figure S5, which illustrates the contributions of distance to nearest instrument, local variability in AQ, and sensor measurement error to large misclassifications of the AQI (off by more than one class) for each of the LCS placement strategies.

## 4. Discussion

In this study, we investigated the utility of including measurements from LCS in real-time AQ reporting using simulations based closely on real data. By comparing different types and amounts of sensor measurement error as well as different LCS placement strategies and numbers of LCS deployed, we were able to differentiate between the impacts of three components of error in daily AQ reporting: distance to the nearest AQ instrument, local variability in AQ, and sensor measurement error.

Our findings offer several key insights and suggestions. One of the most important is that *the value of using LCS for real-time AQ reporting depends strongly on the amount and type of sensor measurement error, and also on the metric to which people pay attention* (i.e., absolute concentration vs. AQI classification). Considering MAE, deploying LCS assuming 10% measurement error (either differential or non-differential) improves daily AQ reporting (compared to only using EPA monitors) across the board. However, among the placement strategies considered here, only LCS placements at schools and PurpleAir locations improve AQ reporting when sensor measurement error is assumed to be 25% and non-differential. Deploying LCS assuming 25% differential measurement error worsens daily AQ reporting across the board.

By contrast, introducing any number of LCS with any amount of sensor measurement error tends to increase the rate of UH misclassifications, which may be even more relevant than absolute $PM_{2.5}$ concentrations for public health (whereas metrics like MAE may be more relevant for AQ-health science). This indicates that *the existing EPA monitor network in California is relatively good at reporting whether the AQ is healthy or unhealthy.* However, the UHM rate begins decreasing as more than 500 LCS are deployed at schools when sensor measurement error is assumed to be 10% or 25% and nondifferential. If sensor measurement error is 10% nondifferential, then deploying LCS at all schools in California (n=7,548) produces a lower rate of UHMs than EPA monitors alone (it is about the same if sensor measurement error is 25% nondifferential). Of course, organizations deploying LCS will need to balance these considerations with their budgets for purchasing and maintaining LCS.

Accounting for both absolute concentration and AQI classification metrics, *it appears that placing LCS at schools results in the most accurate and equitable distribution of daily AQ information* when sensor measurement error is assumed to be less than 25% and nondifferential and more than 500 LCS are deployed. The latter is quite realistic given that in California in 2021, there were 4,343 1 km x 1 km grids with PurpleAir, not to mention other brands of LCS. From a health standpoint, children's relatively high vulnerability to air pollution [41] further motivates the strategy of placing LCS at schools.

For our empirically-based simulation, the degree of error injected into our simulated LCS measurements was drawn from the empirical distribution of errors between collocated California

S15

EPA monitor and PurpleAir sensor data after correction using a national equation developed by EPA researchers [36]. This degree of measurement error is believed to most closely reflect error in the publicly available LCS measurements available through many programs and platforms today. Our simulations show that, under these conditions, using data from LCS in real-time AQ reporting would worsen the accuracy of information from people's nearest AQ instrument, both for the overall population and for the marginalized groups considered. This result suggests that *more region-specific LCS calibration procedures may be necessary for this application of LCS*, which aligns with the findings of several recent studies advocating for region-specific corrections [21,42].

When sensor measurement error is assumed to be high, LCS placement strategies that prioritize those burdened by environmental pollution and sociodemographic injustice result in the most equitable provision of AQ information. Also, across all levels of sensor measurement error and numbers of LCS deployed, selecting LCS placements based on CES Score results in the lowest rate of UHMs for CBGs with high % poverty. Lastly, our simulations revealed that AQ information accuracy under the CES and Pollution Score-based placements is often affected by high local variability in air quality, which makes sense because these locations tend to be near major sources of air pollution. This indicates that *while placing LCS in environmental justice hotspots may benefit those in the immediate community, integrating their data into wider AQ reporting platforms may lead to worsening of real-time AQ information for those outside the immediate community.*

Balancing policy priorities related to LCS deployment will not be easy. To balance the needs of people in less densely populated communities with those of people in more densely populated communities, our analysis unweighted by population density (results shown in the SI) suggests that *some strategic deployments near major roads (especially in less densely populated areas) might also be beneficial*.

These results can inform future investment in LCS networks for equitable AQ monitoring programs in the US, and the methods used can inform similar studies in other locales. While our findings are based on PurpleAir sensors for $PM_{2.5}$, this work informs accuracy targets and larger concerns about LCS placement and calibration across brands of $PM_{2.5}$ sensors, and possibly for other air pollutants. However, it is important to note that this analysis has focused on the use of LCS for provision of real-time AQ information to the public. *These insights do not necessarily transfer to other applications of LCS.* For example, when used for research purposes, LCS have been shown to help capture neighborhood-scale $PM_{2.5}$ when fused with satellite data [43] or incorporated into either a kriging model [44–46] or a machine learning model with spatially-varying correction for the LCS [24]. That said, *the simulation methodology developed for this study could be adapted for many other research questions.*

4.1 Limitations and Future Directions

Design-related limitations of our study are that we only used data from California and that we assumed individuals view the AQ information from the instrument nearest to their location of residence as their personal exposure. A technical limitation of this analysis is our use of the Di et al. daily 1 km x 1 km estimates for the "true" exposures. AQ may vary substantially over 24 hours and over a 1 km x 1 km square, and as several studies have observed, the $PM_{2.5}$ patterns sensed by LCS are often different (i.e. affected more by local sources) than EPA monitors [47,48], which Di et al. used to train their model. We also did not consider any differences in LCS performance due to varying PM composition, which have been observed elsewhere [35].

Future work might consider more nuance in the LCS measurement error problem, such as accounting for sensor "drift" [20,49,50] and varying particle composition / meteorological conditions, as has been explored by EPA researchers who have proposed a different correction method for wildfire smoke measured by PurpleAir sensors [51]. This might be addressed by considering more spatial and temporal correlation in the LCS measurement errors. In terms of LCS placement, while we chose relatively simple selection strategies to facilitate comprehension and comparison, future research could harness more sophisticated statistical and/or atmospheric modeling techniques [45,52,53] to identify locations yielding key spatiotemporal information or to prioritize some LCS placements near EPA monitors for the purposes of sensor calibration. Finally, there is more work to be done investigating how access to real-time AQ information and/or alerts translates into public health and economic benefits, as several studies have begun exploring [54–58].

**Supporting Information** (see second half of this document)
- Supplemental notes on LCS device measurement error simulation.
- Supplemental figures and tables for all analyses: maps visualizing LCS placement strategies and distributions of marginalized groups, descriptions of contextual datasets and processing steps, annual average $PM_{2.5}$ summaries and other descriptive statistics, hypothetical numbers of LCS in well-known counties under each placement strategy, distributions of simulated and empirical sensor measurement error, and summary statistics of the Di et al. estimates as used for the empirical sensor measurement error simulation.
- Supplemental figures and tables for the analysis weighted by population density: basic descriptive statistics, distance to the nearest AQ instrument among observations misclassified by more than one level of the AQI, underclassifications and overclassifications of the AQI, metrics with LCS deployed at all schools (n = 7,548).
- Supplemental figures and tables for the analysis unweighted by population density: counterparts of all figures and tables (from the weighted analysis) in the main text and SI.

**Acknowledgments**

We thank the National Studies on Air Pollution and Health (NSAPH) research lab, specifically the biostatistics working group led by Francesca Dominici, for their support of this project.

The computations in this paper were run on the FASRC Cannon cluster supported by the FAS Division of Science Research Computing Group at Harvard University.

**Funding:**
National Institutes of Health grant 5T32ES007142 (EMC)
National Institutes of Health grant 1K01ES032458 (RCN)

**Author contributions:**
  Conceptualization: EMC, RCN, PD, DB
  Methodology: EMC, RCN, PD, DB
  Investigation: EMC
  Visualization: EMC
  Supervision: RCN, PD, DB
  Writing—original draft: EMC
  Writing—review & editing: EMC, LK, RCN, PD, DB

**Competing interests:** Authors declare that they have no competing interests.

**Data and materials availability:**
Code used to download and process the datasets, run the analyses, and generate the figures and tables can be found at https://github.com/EllenConsidine/LCS_placement_sims. All data used in this study are publicly available and our analytic (processed) dataset is on Harvard Dataverse, at https://doi.org/10.7910/DVN/QR4N7V. Descriptions of the data sources are in the Materials & Methods section as well as our GitHub README file.

# Supporting Information for

# Investigating Use of Low-Cost Sensors to Increase Accuracy and Equity of Real-Time Air Quality Information

Ellen M. Considine*, Danielle Braun, Leila Kamareddine, Rachel C. Nethery, and Priyanka deSouza

*Corresponding author. Email: ellen_considine@g.harvard.edu

**Contents (27 pages):**

**Supplemental Notes on LCS Device Measurement Error Simulation**

*This section accompanies subsection 2.4 in the main text.*

Generating Errors Based on Proposed Performance Targets

As a benchmark, we first ran simulations using no sensor measurement error (as if the LCS were as accurate as the EPA monitors). Then, we used two levels of both non-differential and differential sensor measurement error. For a given accuracy level (either ±10% or ±25%, as proposed in the EPA performance targets workshop [35]), we simulated non-differential sensor measurement error by sampling from a normal distribution with mean zero and standard deviation of 5 (the unweighted 2016 average $PM_{2.5}$ in California) multiplied by the accuracy level, yielding 0.5 and 1.25 respectively. In mathematical notation:

$\{LCS\ PM_{2.5}\} = \{Di\ et\ al.\ PM_{2.5}\} + \varepsilon_n$

where $\varepsilon_n \sim Normal(mean = 0, sd = \{0.1\ or\ 0.25\}*5)$

The differential sensor measurement error, on the other hand, was motivated by our observation of the positive correlation between the magnitude of the EPA correction residuals and the true $PM_{2.5}$ (described in the following subsection). For the two accuracy levels, we simulated differential sensor measurement error by sampling from a normal distribution with mean zero and standard deviation of the true $PM_{2.5}$ (for each grid and day) multiplied by the accuracy level. In mathematical notation:

$\{LCS\ PM_{2.5}\} = \{Di\ et\ al.\ PM_{2.5}\} + \varepsilon_d$

where $\varepsilon_d \sim Normal(mean = 0, sd = \{0.1\ or\ 0.25\}*\{Di\ et\ al.\ PM_{2.5}\})$

Sampling from the Empirical Distribution of PurpleAir Errors

To obtain an empirical distribution of measurement errors for our simulated LCS, we compared PurpleAir measurements and EPA monitor measurements for collocated EPA monitor and LCS pairs in California. Note that these data were not used directly in the simulations: the estimates from Di et al. were considered to be the true exposure. We obtained daily $PM_{2.5}$ averages for both the monitors and LCS from the year 2020 (as opposed to 2016, the year for which we had 1 km x 1 km estimates from Di et al. to use as "truth" in the simulations) because most of the PurpleAir sensors have been deployed since 2016. We used the purpleair Python package to obtain measurements of $PM_{2.5}$, temperature, and relative humidity from PurpleAir sensors (channel A on 1/11/22 and channel B on 1/23/22), and downloaded the EPA monitor measurements from the EPA Air Quality System annual summary files repository [59].

Following the 24-hour quality assurance methods of the EPA researchers who developed a national correction for PurpleAir data [36], we identified all pairs of EPA monitors and outdoor PurpleAir sensors within 50 meters of each other. Then, also following their quality assurance methods, we only considered EPA monitor averages based on at least 18 hours of observation, and PurpleAir averages based on at least 90% of the 2-minute observations, where the difference in daily averages of the two PurpleAir channels was less than 5 $\mu g/m^3$ or 61%. Then, we averaged the two PurpleAir channels. This procedure yielded 5,759 paired daily observations from 22 EPA monitor-PurpleAir locations. To make this distribution more realistic for 2016, we removed 48 daily observations where the EPA monitor $PM_{2.5}$ was greater than 112 $\mu g/m^3$ (the maximum value in the 2016 Di et al. estimates), yielding n = 5,711.

The EPA correction equation developed for PurpleAir sensors in the US is

$$PM_{2.5} = 0.524 \times \{PurpleAir\ PM_{2.5}\} - 0.0862 \times \{Relative\ Humidity\} + 5.75$$

After applying this correction to the PurpleAir data, we compared the corrected values to the collocated EPA monitor measurements. The resulting RMSE (root-mean squared error) was $5.33 \mu g/m^3$ and $R^2$ was 0.81. We refer to the difference between these EPA monitor and corrected LCS measurements as the empirical residuals.

Figure S4 shows that the magnitude of empirical residuals increased as the true $PM_{2.5}$ concentration (as measured by the EPA monitors) increased. Sensor measurement error increasing with $PM_{2.5}$ concentration (i.e., differential measurement error) has been observed by some other groups as well [18]. Such a relationship was not observed for relative humidity nor ambient temperature. Thus, for the empirically based LCS measurement error simulation, we sampled from the distribution of residuals corresponding to the decile of true $PM_{2.5}$ (the Di et al. estimates) at each grid and day. A summary of the Di et al. estimates mapped into these deciles is given in Table S3.

**Supplemental Figures and Tables for All Analyses**

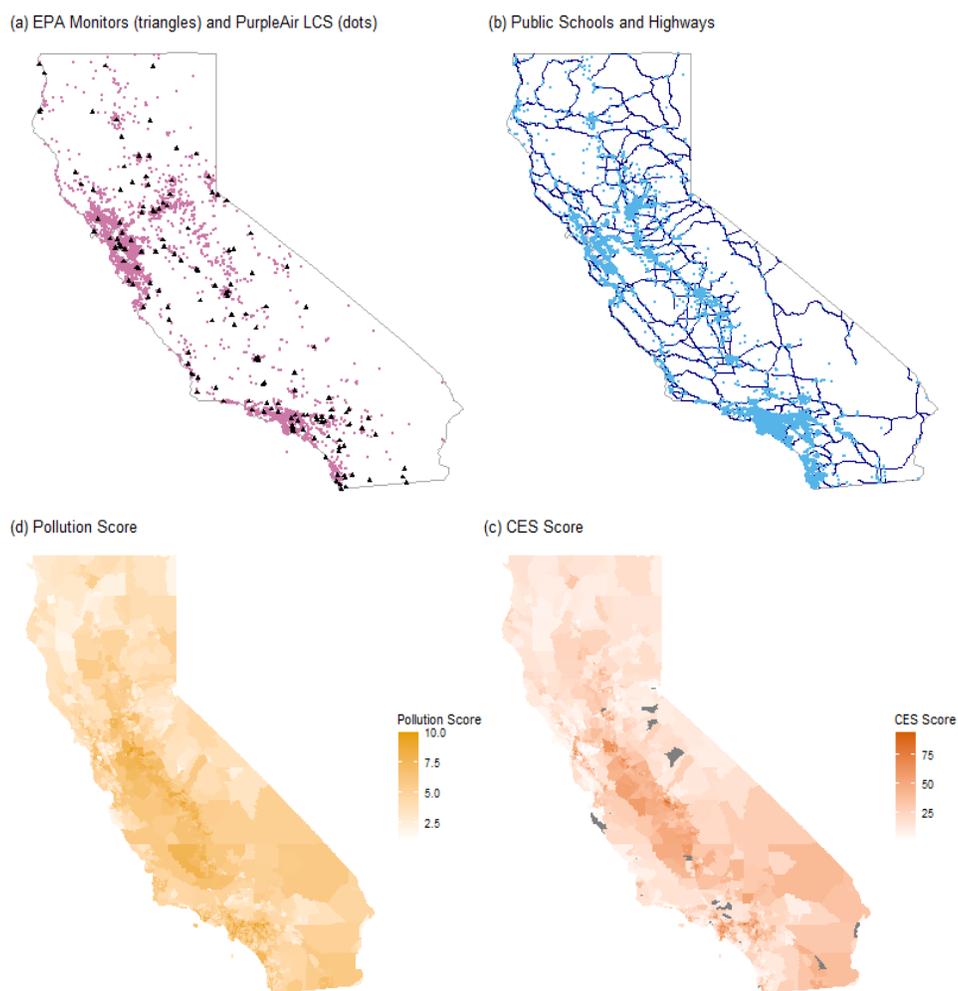

**Fig. S1. Contextual Maps.** Visualizing data used to guide the hypothetical LCS placement strategies. Locations of EPA monitors and PurpleAir LCS (panel a), public schools and major roads in California (panel b), as well as maps of CES (panel c) and Pollution Score (panel d). Data sources (all public domain / open access): EPA AQS and PurpleAir (panel a), NCES and NHPN (panel b), California EPA's OEHHA (panels c and d).

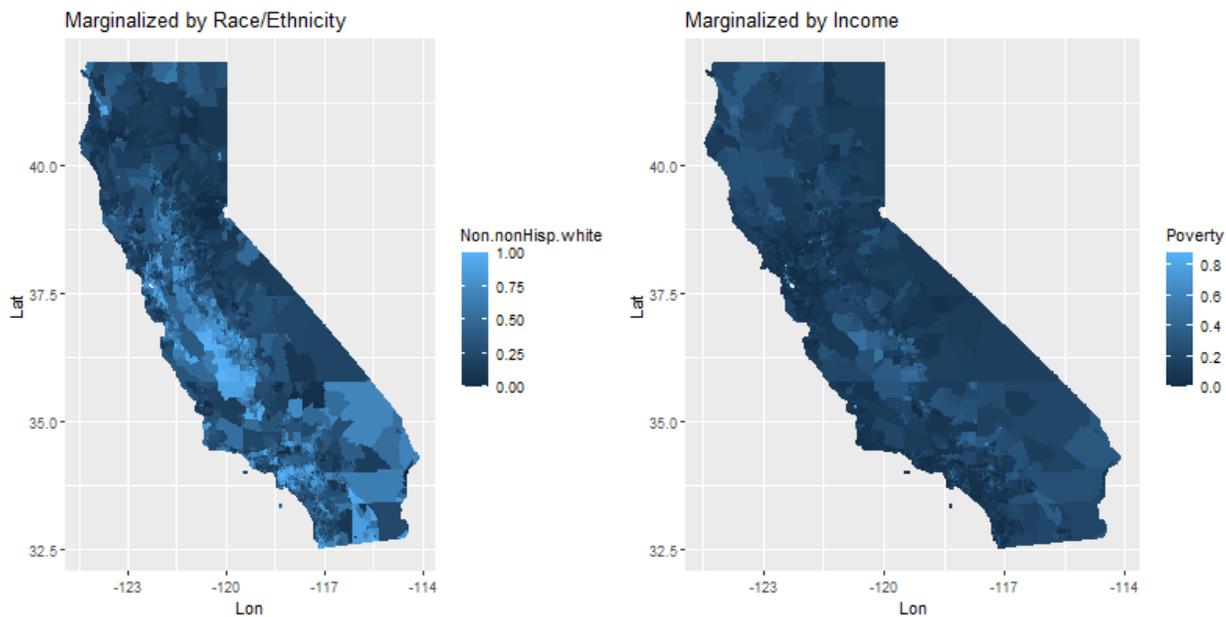

**Fig. S2. Distributions of marginalized groups** considered in the analyses: percent of individuals identifying as non-white or Hispanic, and percent of households with income below the poverty line. Data source: 2016 ACS (public domain).

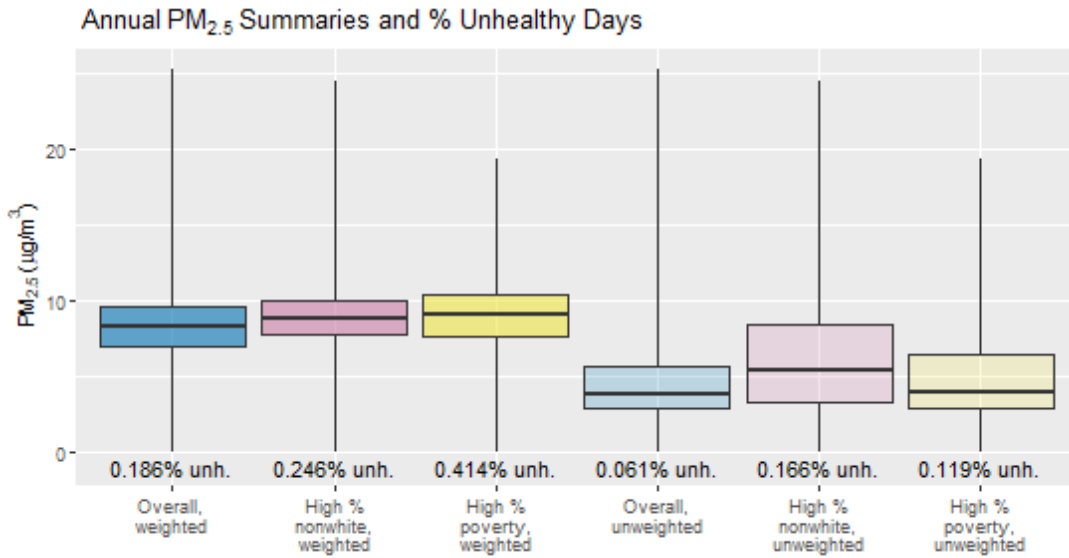

**Fig. S3. Annual average PM₂.₅ ($\mu$g/m$^3$) summaries** for population subsets, weighted and unweighted by population density; labeled with % Unhealthy (AQI classification = Orange and higher) days.

**Table S1. Descriptions of Contextual Datasets.** Describing the data sets, data processing steps, and sampling methods used to select hypothetical locations for LCS in each simulation.

| Placement strategy | Data source | Data processing | LCS location selection method (within each simulation) | Additional notes |
|---|---|---|---|---|
| Current PurpleAir Locations | The purpleair Python package [60] | Assigned each outdoor PurpleAir sensor to its nearest 1 km x 1 km grid centroid | Randomly selected grids with PurpleAir | 4,343 grids with PurpleAir LCS |
| Public School Locations | The National Center for Education Statistics [61] | Assigned each school to its nearest 1 km x 1 km grid centroid | Randomly selected grids with schools | 7,548 grids with schools |
| Near-Road | The National Highway Planning Network shapefile [62] | Summed the lengths of major roads within circular buffers with 500 m radii around each grid centroid | Randomly sampled from all grids, where the sampling probability for each grid was the sum of road lengths | 4.5% of the grids had major roads; among these, the average length was 2km |
| CES Score | Cal Enviro Screen 3.0 [34] | Assigned all grids the CES Score of the census tract their centroid falls within | Randomly sampled from all grids, where the sampling probability for each grid was the CES Score | *Additional notes are in the main text* |
| Pollution Score | Cal Enviro Screen 3.0 [34] | Assigned all grids the Pollution Score of the census tract their centroid falls within | Randomly sampled from all grids, where the sampling probability for each grid was the Pollution Score | *Additional notes are in the main text* |

**Table S2. Hypothetical numbers of LCS in well-known counties.** Average number of LCS (out of 1,000 deployments) placed in the counties of Sacramento, Imperial, and Los Angeles according to each placement strategy considered in this study.

| County | Population size (2016) | Avg. CES Score | Purple Air (current) | Schools | Near road locations | CES Score based | Pollution Score based |
|---|---|---|---|---|---|---|---|
| Sacramento | 1,495,611 | 23 | 26 | 40 | 10 | 7 | 7 |
| Imperial | 186,019 | 38 | 1 | 7 | 26 | 46 | 34 |
| Los Angeles | 10,180,169 | 23 | 107 | 204 | 30 | 25 | 26 |

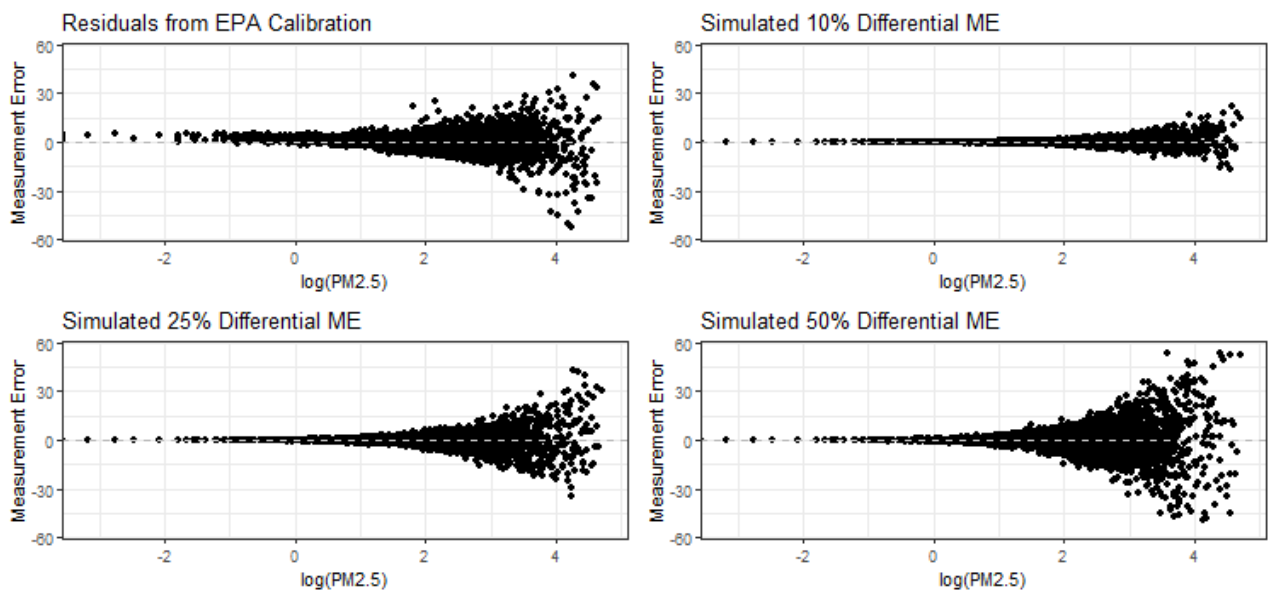

**Fig. S4. Simulated and empirical sensor measurement error.** Top left plot shows the empirical distribution of residuals in $\mu g/m^3$ (compared to AQS monitor measurements within 50m) from PurpleAir observations corrected using the EPA equation. The other three plots show the distributions resulting from our method of simulating sensor measurement error differentially with respect to true $PM_{2.5}$ concentration. Note that 50% accuracy was not included in the full analysis simulations.

**Table S3. Summary statistics of the Di et al. estimates** in each decile of AQS PM$_{2.5}$ from our correction dataset, in $\mu$g/m$^3$. Matching on these deciles was used to draw empirical residuals from the EPA correction applied to collocated PurpleAir and AQS data.

| Decile | Minimum | Q1 | Median | Mean | Q3 | Maximum |
|---|---|---|---|---|---|---|
| 1 | 0.000 | 0.684 | 1.265 | 1.262 | 1.863 | 2.455 |
| 2 | 2.455 | 2.757 | 3.074 | 3.083 | 3.404 | 3.750 |
| 3 | 3.750 | 4.055 | 4.375 | 4.389 | 4.716 | 5.083 |
| 4 | 5.083 | 5.385 | 5.711 | 5.731 | 6.066 | 6.458 |
| 5 | 6.458 | 6.780 | 7.136 | 7.167 | 7.538 | 8.000 |
| 6 | 8.000 | 8.365 | 8.778 | 8.818 | 9.251 | 9.796 |
| 7 | 9.796 | 10.225 | 10.705 | 10.747 | 11.246 | 11.875 |
| 8 | 11.88 | 12.45 | 13.15 | 13.25 | 13.99 | 15.08 |
| 9 | 15.08 | 15.93 | 17.10 | 17.60 | 18.91 | 22.65 |
| 10 | 22.65 | 24.24 | 26.57 | 29.49 | 31.02 | 112.18 |

**Supplementary Figures and Tables for the Analysis Weighted by Population Density**

**Table S4: Basic descriptive statistics, weighted by population density** (unless otherwise specified) and normalized, to provide a sense of scale. Non-normalized results are in Table S5. All cells are of the form "Mean (Standard Deviation)". Note: wtd. = weighted. The rows that specify when population density greater than 500 / sq. mile is to account for both the weighting to select the LCS locations (i.e., by proximity to roads, Pollution Score, or CES Score) and the weighting by population density in our results. Population density less than 500 / sq. mile is considered rural by the USDA [63].

| Locations | Annual Avg. PM$_{2.5}$ ($\mu$g/m$^3$) | % Poverty | CES Score | % Nonwhite or Hispanic | Population Density (unweighted) |
|---|---|---|---|---|---|
| CA overall | 0 (1) | 0 (1) | 0 (1) | 0 (1) | 0 (1) |
| EPA monitor sites | 0.34 (1.01) | 0.67 (1) | 0.62 (1.06) | 0.22 (0.96) | 2.54 (3.99) |
| PurpleAir sites | -0.24 (0.83) | -0.33 (0.83) | -0.46 (0.9) | -0.26 (0.93) | 2.51 (4.11) |
| School sites | 0.23 (0.89) | 0.17 (1) | 0.23 (1.02) | 0.33 (0.93) | 3.96 (4.28) |
| Favoring by nearby roads (not wtd. by pop. density) | -1.37 (1.23) | 0.08 (0.75) | -0.29 (0.81) | -0.93 (0.96) | 0.3 (1.75) |
| Favoring by roads, where pop. density > 500 | -0.28 (1.06) | -0.08 (1) | -0.16 (0.97) | -0.3 (1) | 2.87 (3.67) |
| Favoring by Pollution Score (not wtd. by pop. density) | -1.7 (1.26) | 0.17 (0.67) | -0.1 (0.79) | -0.81 (0.93) | 0.02 (1.07) |
| Favoring by Pollution Score, where pop. density > 500 | 0 (1.03) | -0.08 (1) | 0.03 (1.04) | -0.15 (0.96) | 2.81 (3.48) |
| Favoring by CES Score (not wtd. by pop. density) | -1.67 (1.31) | 0.25 (0.75) | 0.11 (0.8) | -0.67 (0.96) | 0.03 (1.16) |
| Favoring by CES Score, where | 0.12 (1.06) | 0.17 (1.08) | 0.41 (1.04) | 0.11 (0.93) | 3.13 (3.79) |

| pop. density > 500 | | | | | |
|---|---|---|---|---|---|
| High % nonwhite | 0.31 (0.9) | 0.33 (1) | 0.49 (0.93) | 0.78 (0.44) | 0.29 (1.83) |
| High % poverty | 0.37 (1.09) | 1.5 (0.75) | 0.92 (0.9) | 0.7 (0.78) | 0.02 (1.26) |

**Table S5: Non-normalized version of Table S4** (above). Basic descriptive statistics, weighted by population density (unless otherwise specified). All cells are of the form "Mean (Standard Deviation)". Note: wtd. = weighted. The rows that specify when population density greater than 500 / sq. mile is to account for both the weighting to select the LCS locations (i.e. by proximity to roads, Pollution Score, or CES Score) and the weighting by population density in our results. Population density less than 500 / sq. mile is considered rural.

| Locations | Annual Avg. $PM_{2.5}$ ($\mu g/m^3$) | % Poverty | CES Score | % Nonwhite or Hispanic | Population Density (unweighted) |
|---|---|---|---|---|---|
| CA overall | 8.23 (1.98) | 0.16 (0.12) | 28.11 (15.93) | 0.61 (0.27) | 249.81 (1,526.04) |
| EPA monitor sites | 8.91 (2) | 0.24 (0.12) | 37.98 (16.95) | 0.67 (0.26) | 4,119.08 (6,090.22) |
| PurpleAir sites | 7.76 (1.65) | 0.12 (0.1) | 20.84 (14.33) | 0.54 (0.25) | 4,075.79 (6,267.25) |
| School sites | 8.69 (1.77) | 0.18 (0.12) | 31.74 (16.28) | 0.70 (0.25) | 6,296.3 (6,531.5) |
| Favoring by nearby roads (not wtd. by pop. density) | 5.52 (2.43) | 0.17 (0.09) | 23.42 (12.96) | 0.36 (0.26) | 701.74 (2,670.86) |
| Favoring by nearby roads, where pop. density > 500 | 7.68 (2.1) | 0.15 (0.12) | 25.63 (15.5) | 0.53 (0.27) | 4,627.07 (5,604.17) |
| Favoring by Pollution Score (not wtd. by pop. density) | 4.87 (2.49) | 0.18 (0.08) | 26.52 (12.51) | 0.39 (0.25) | 275.26 (1,633.18) |
| Favoring by Pollution Score, where pop. density > 500 | 8.23 (2.04) | 0.15 (0.12) | 28.51 (16.57) | 0.57 (0.26) | 4,535.05 (5,309.65) |
| Favoring by CES Score (not wtd. by pop. density) | 4.93 (2.6) | 0.19 (0.09) | 29.88 (12.71) | 0.43 (0.26) | 296.21 (1,774.24) |
| Favoring by CES Score, where | 8.46 (2.09) | 0.18 (0.13) | 34.68 (16.54) | 0.64 (0.25) | 5,027.12 (5,779.83) |

| pop. density > 500 | | | | | |
|---|---|---|---|---|---|
| High % nonwhite | 8.84 (1.78) | 0.20 (0.12) | 35.92 (14.87) | 0.82 (0.12) | 696.8 (2,789.15) |
| High % poverty | 8.96 (2.16) | 0.34 (0.09) | 42.81 (14.28) | 0.80 (0.21) | 277.93 (1,922) |

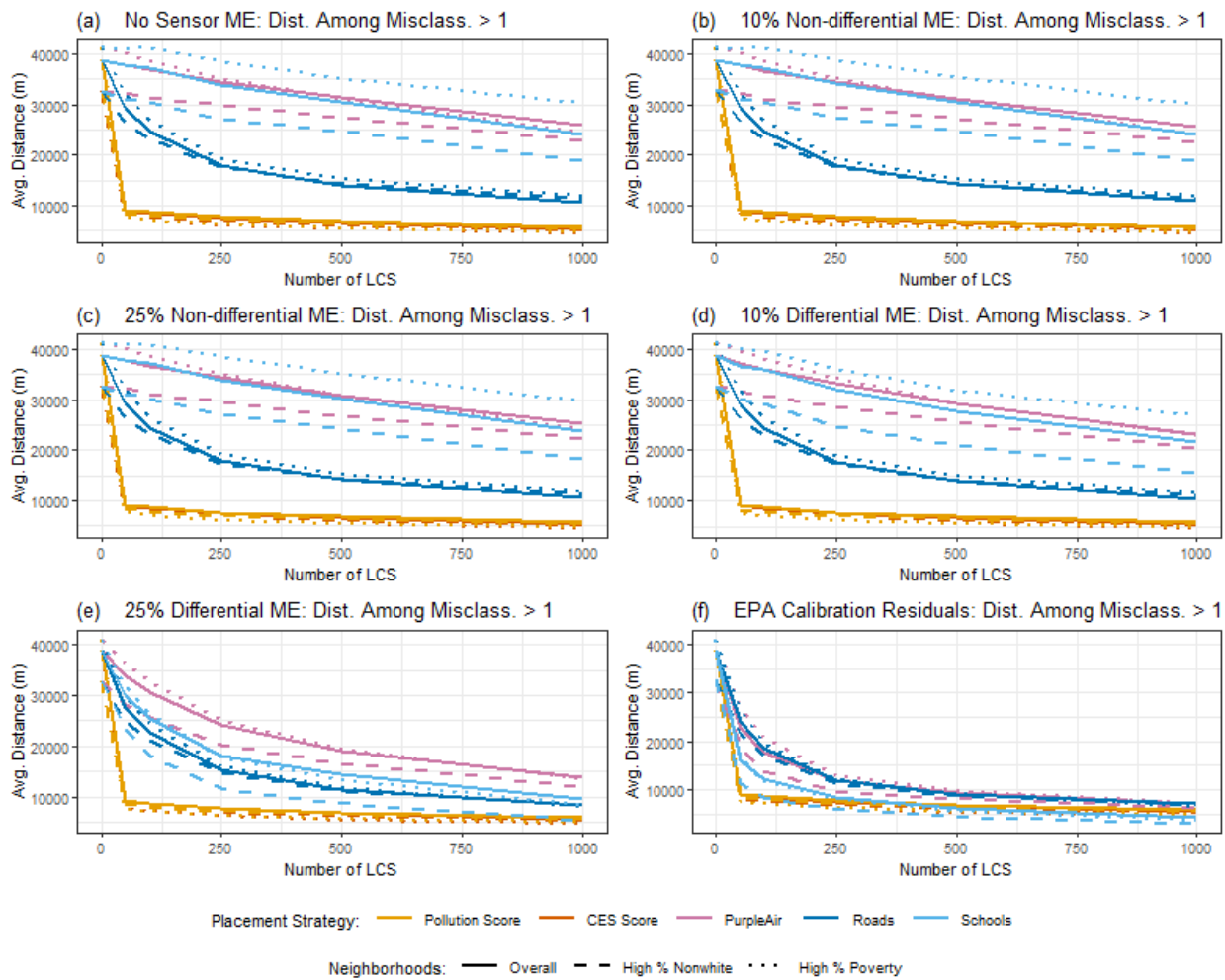

**Fig. S5. Distance to the nearest AQ instrument, among observations misclassified by more than one level of the AQI** (e.g. Red to Yellow, or Green to Orange). All results were calculated using 366 days and averaged across 100 simulation replicates, weighted by population density.

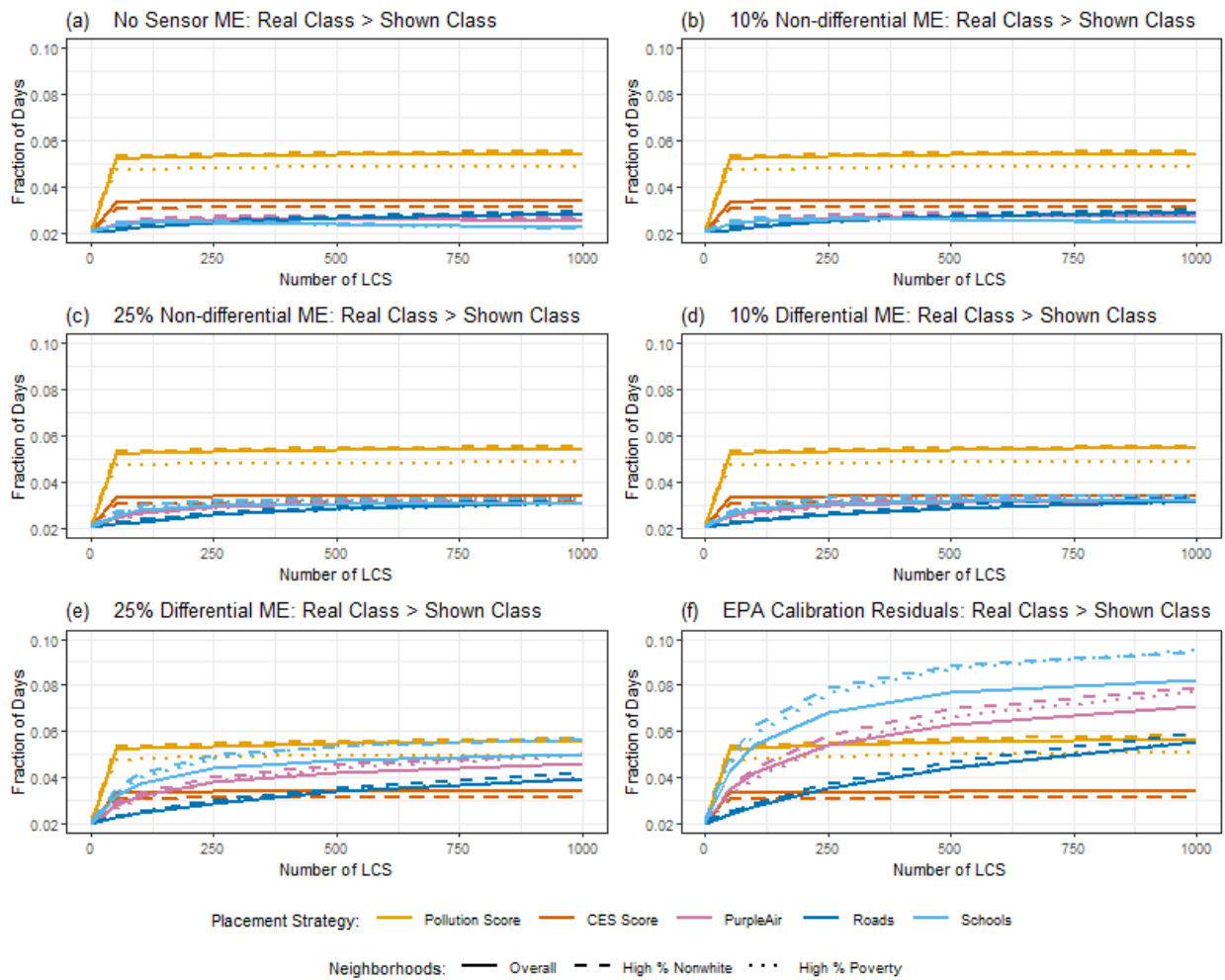

**Fig. S6. Underclassifications of the AQI.** The average fraction of days on which the real AQI class (what was truly experienced) was higher than the shown AQI class (reported from the nearest monitor or sensor), resulting from different numbers of LCS deployed, LCS placement strategies, and sensor measurement error types and amounts. All results were calculated using 366 days and averaged across 100 simulation replicates, weighted by population density.

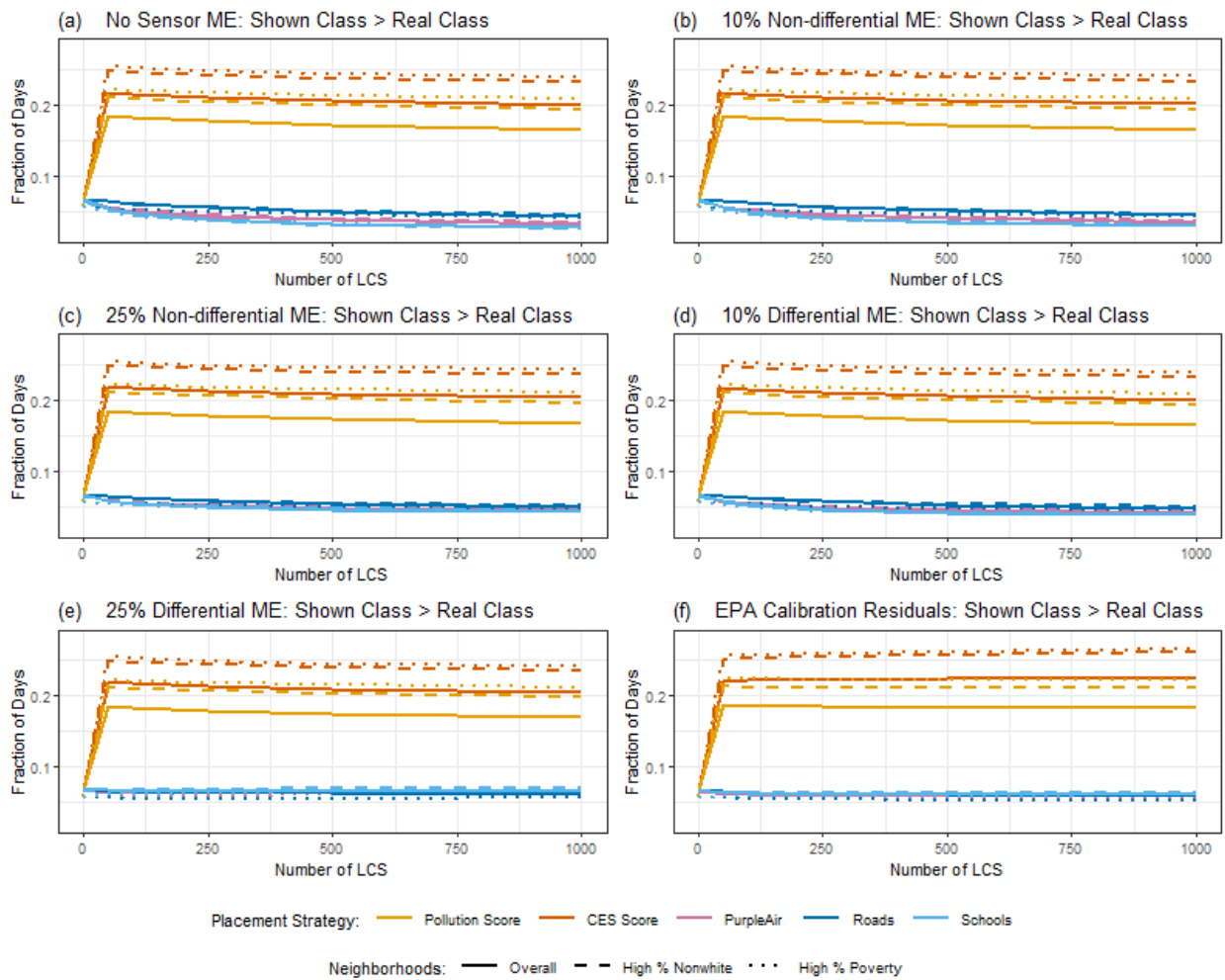

**Fig. S7. Overclassifications of the AQI.** The average fraction of days on which the shown AQI class (reported from the nearest monitor or sensor) was higher than the real AQI class (what was truly experienced), resulting from different numbers of LCS deployed, LCS placement strategies, and sensor measurement error types and amounts. All results were calculated using 366 days and averaged across 100 simulation replicates, weighted by population density.

**Table S6: Comparing impacts of different sensor accuracies for LCS at all schools** (similar to Table 1 in the main text), weighted by population density. Results when there are no LCS vs. LCS at all school locations (n = 7,548), assuming different kinds and amounts of sensor measurement error. Results are the average of 100 simulation replicates to account for randomness in the sensor measurement error generation. Unless otherwise specified, "errors" refer to the difference between the true exposure experienced at each grid centroid and the exposure reported from the nearest AQ instrument. AQI under-classification is when the true exposure class is greater than what someone is shown, and over-classification is when the true exposure class is less than what someone is shown. Rate of UH misclassification (UHM) is the fraction of days with unhealthy AQI (Orange+) that are misreported as healthy AQI (Green or Yellow).

| Type/Amount of Sensor Measurement Error | Std. Dev. of Sensor Measurement Error ($\mu g/m^3$) | MAE ($\mu g/m^3$) | 95th Percentile of Errors ($\mu g/m^3$) | Under-classified AQI (%) | Over-classified AQI (%) | UHM (%) |
|---|---|---|---|---|---|---|
| Overall Population | | | | | | |
| No LCS (only EPA monitors) | — | 1.46 | 4.45 | 2.05 | 6.79 | 11.37 |
| No Sensor Error | 0 | 0.53 | 2.30 | 1.32 | 1.48 | 8.40 |
| 10% Non-differential | 0.5 | 0.76 | 2.42 | 1.88 | 2.18 | 9.74 |
| 25% Non-differential | 1.25 | 1.24 | 3.20 | 2.79 | 3.80 | 11.82 |
| 10% Differential | 0.99 | 0.97 | 2.92 | 2.90 | 3.23 | 18.41 |
| 25% Differential | 2.46 | 1.84 | 5.31 | 5.10 | 6.48 | 29.22 |
| EPA Correction Residual Decile Draws | 3.70 | 2.51 | 7.22 | 8.72 | 6.07 | 28.14 |
| Population living in CBGs with high % nonwhite | | | | | | |
| No LCS (only EPA monitors) | — | 1.34 | 4.14 | 2.15 | 6.53 | 10.35 |
| No Sensor Error | 0 | 0.39 | 1.81 | 1.11 | 1.25 | 7.22 |
| 10% Non-differential | 0.5 | 0.65 | 1.97 | 1.81 | 2.11 | 8.71 |
| 25% Non-differential | 1.25 | 1.16 | 2.94 | 2.93 | 3.95 | 11.09 |
| 10% Differential | 1.03 | 0.91 | 2.69 | 3.06 | 3.34 | 18.32 |
| 25% Differential | 2.56 | 1.88 | 5.42 | 5.75 | 6.92 | 29.44 |
| EPA Correction Residual Decile Draws | 3.83 | 2.59 | 7.49 | 10.03 | 6.39 | 28.50 |
| Population living in CBGs with high % poverty | | | | | | |
| No LCS (only EPA monitors) | — | 1.28 | 4.06 | 2.09 | 5.79 | 8.40 |
| No Sensor Error | 0 | 0.42 | 1.97 | 1.14 | 1.22 | 5.15 |
| 10% Non-differential | 0.5 | 0.68 | 2.11 | 1.87 | 2.08 | 6.55 |
| 25% Non-differential | 1.25 | 1.18 | 3.01 | 3.03 | 3.85 | 8.75 |
| 10% Differential | 1.08 | 0.95 | 2.83 | 3.17 | 3.29 | 15.88 |
| 25% Differential | 2.71 | 1.93 | 5.60 | 6.00 | 6.72 | 27.52 |
| EPA Correction Residual Decile Draws | 4.03 | 2.64 | 7.66 | 10.42 | 6.22 | 26.47 |

**Supplementary Figures and Tables for the Analysis Not Weighted by Population Density**

**Table S7: Basic descriptive statistics, unweighted by population density** (counterpart of Table S5). All cells are of the form "Mean (Standard Deviation)". Note: wtd. = weighted.

| Locations | Annual Avg. PM$_{2.5}$ ($\mu$g/m$^3$) | % Poverty | CES Score | % Nonwhite or Hispanic | Population Density |
|---|---|---|---|---|---|
| CA overall | 4.55 (2.32) | 0.17 (0.08) | 23.92 (11.95) | 0.35 (0.24) | 249.81 (1,526.04) |
| EPA monitor sites | 7.43 (2.58) | 0.20 (0.13) | 29.23 (15.8) | 0.53 (0.26) | 4,119.08 (6,090.22) |
| PurpleAir sites | 6.63 (2.05) | 0.11 (0.09) | 16.47 (12.32) | 0.38 (0.24) | 4,075.79 (6,267.25) |
| School sites | 8.04 (2.08) | 0.16 (0.11) | 27.88 (16.04) | 0.58 (0.27) | 6,296.3 (6,531.5) |
| Favoring by nearby roads | 5.52 (2.43) | 0.17 (0.09) | 23.42 (12.96) | 0.36 (0.26) | 701.74 (2,670.86) |
| Favoring by Pollution Score | 4.87 (2.49) | 0.18 (0.08) | 26.52 (12.51) | 0.39 (0.25) | 275.26 (1,633.18) |
| Favoring by CES Score | 4.93 (2.6) | 0.19 (0.09) | 29.88 (12.71) | 0.43 (0.26) | 296.21 (1,774.24) |
| High % nonwhite | 5.77 (2.93) | 0.24 (0.09) | 38.31 (10.84) | 0.74 (0.12) | 696.8 (2,789.15) |
| High % poverty | 4.85 (2.66) | 0.30 (0.06) | 31.6 (13.5) | 0.52 (0.26) | 277.93 (1,922) |

**Table S8: Comparing impacts of different sensor accuracies, unweighted by population density** (counterpart of Table 1 in the main text). Results when there are no LCS vs. LCS at all real PurpleAir locations (n = 4,343), assuming different kinds and amounts of sensor measurement error. Results are the average of 100 simulation replicates to account for randomness in the sensor measurement error generation. Unless otherwise specified, "errors" refer to the difference between the true exposure experienced at each grid centroid and the exposure reported from the nearest AQ instrument. AQI under-classification is when the true exposure class is greater than what someone is shown, and over-classification is when the true exposure class is less than what someone is shown. Rate of UH misclassification (UHM) is the fraction of days with unhealthy AQI (Orange+) that are misreported as healthy AQI (Green or Yellow).

| Type/Amount of Sensor Measurement Error | Std. Dev. of Sensor Measurement Error ($\mu g/m^3$) | MAE ($\mu g/m^3$) | 95th Percentile of Errors ($\mu g/m^3$) | Under-classified AQI (%) | Over-classified AQI (%) | UHM (%) |
|---|---|---|---|---|---|---|
| Overall Population | | | | | | |
| No LCS (only EPA monitors) | — | 2.51 | 8.08 | 1.17 | 7.09 | 17.67 |
| No Sensor Error | 0 | 1.79 | 5.87 | 1.36 | 3.59 | 25.95 |
| 10% Non-differential | 0.5 | 1.85 | 5.92 | 1.39 | 3.66 | 25.50 |
| 25% Non-differential | 1.25 | 2.08 | 6.19 | 1.47 | 4.09 | 25.42 |
| 10% Differential | 0.78 | 1.86 | 5.99 | 1.51 | 3.84 | 26.88 |
| 25% Differential | 1.95 | 2.12 | 6.68 | 1.85 | 4.86 | 32.45 |
| EPA Correction Residual Decile Draws | 3.06 | 2.68 | 7.34 | 2.62 | 4.61 | 31.19 |
| Population living in CBGs with high % nonwhite | | | | | | |
| No LCS (only EPA monitors) | — | 2.81 | 9.43 | 1.58 | 10.56 | 9.86 |
| No Sensor Error | 0 | 1.79 | 5.60 | 2.11 | 3.76 | 21.41 |
| 10% Non-differential | 0.5 | 1.84 | 5.65 | 2.19 | 3.86 | 20.69 |
| 25% Non-differential | 1.25 | 2.07 | 5.95 | 2.37 | 4.43 | 20.86 |
| 10% Differential | 0.91 | 1.88 | 5.75 | 2.44 | 4.14 | 22.84 |
| 25% Differential | 2.27 | 2.23 | 6.65 | 3.12 | 5.51 | 29.00 |
| EPA Correction Residual Decile Draws | 3.43 | 2.72 | 7.56 | 4.50 | 5.26 | 27.94 |
| Population living in CBGs with high % poverty | | | | | | |
| No LCS (only EPA monitors) | — | 2.95 | 9.40 | 1.11 | 9.63 | 9.89 |
| No Sensor Error | 0 | 1.83 | 5.88 | 1.58 | 3.11 | 16.82 |
| 10% Non-differential | 0.5 | 1.89 | 5.93 | 1.63 | 3.17 | 15.73 |
| 25% Non-differential | 1.25 | 2.11 | 6.21 | 1.73 | 3.64 | 15.97 |
| 10% Differential | 0.88 | 1.90 | 6.00 | 1.78 | 3.39 | 18.04 |
| 25% Differential | 2.20 | 2.18 | 6.70 | 2.20 | 4.56 | 24.47 |
| EPA Correction Residual Decile Draws | 3.38 | 2.70 | 7.38 | 3.11 | 4.42 | 23.74 |

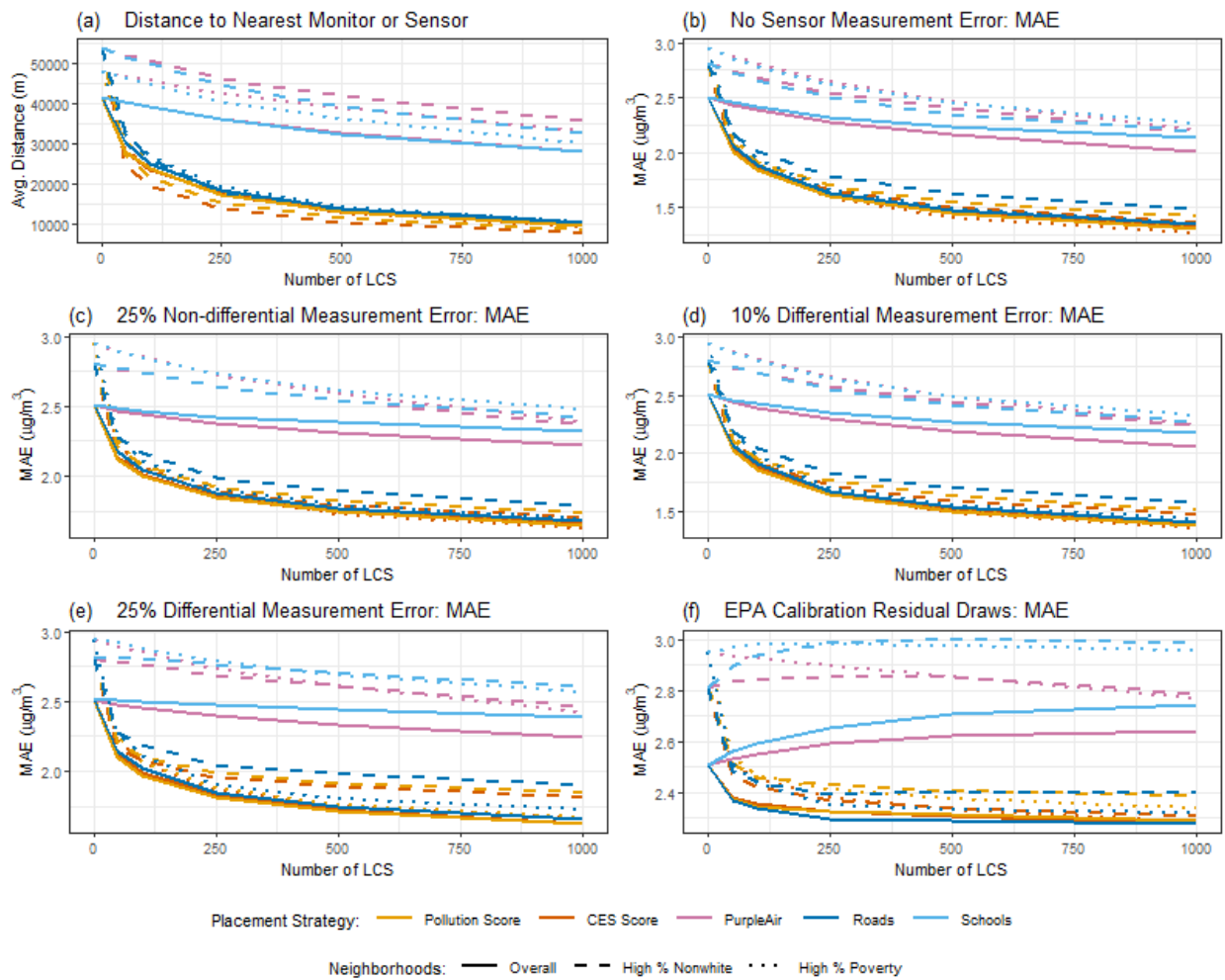

**Fig. S8. Distance and MAE, unweighted by population density.** Distance to the nearest monitor or sensor and mean absolute error (between what is reported vs. experienced) resulting from different numbers of LCS deployed, LCS placement strategies, and sensor measurement error types and amounts. All results were calculated using 366 days and averaged across 100 simulation replicates, unweighted by population density.

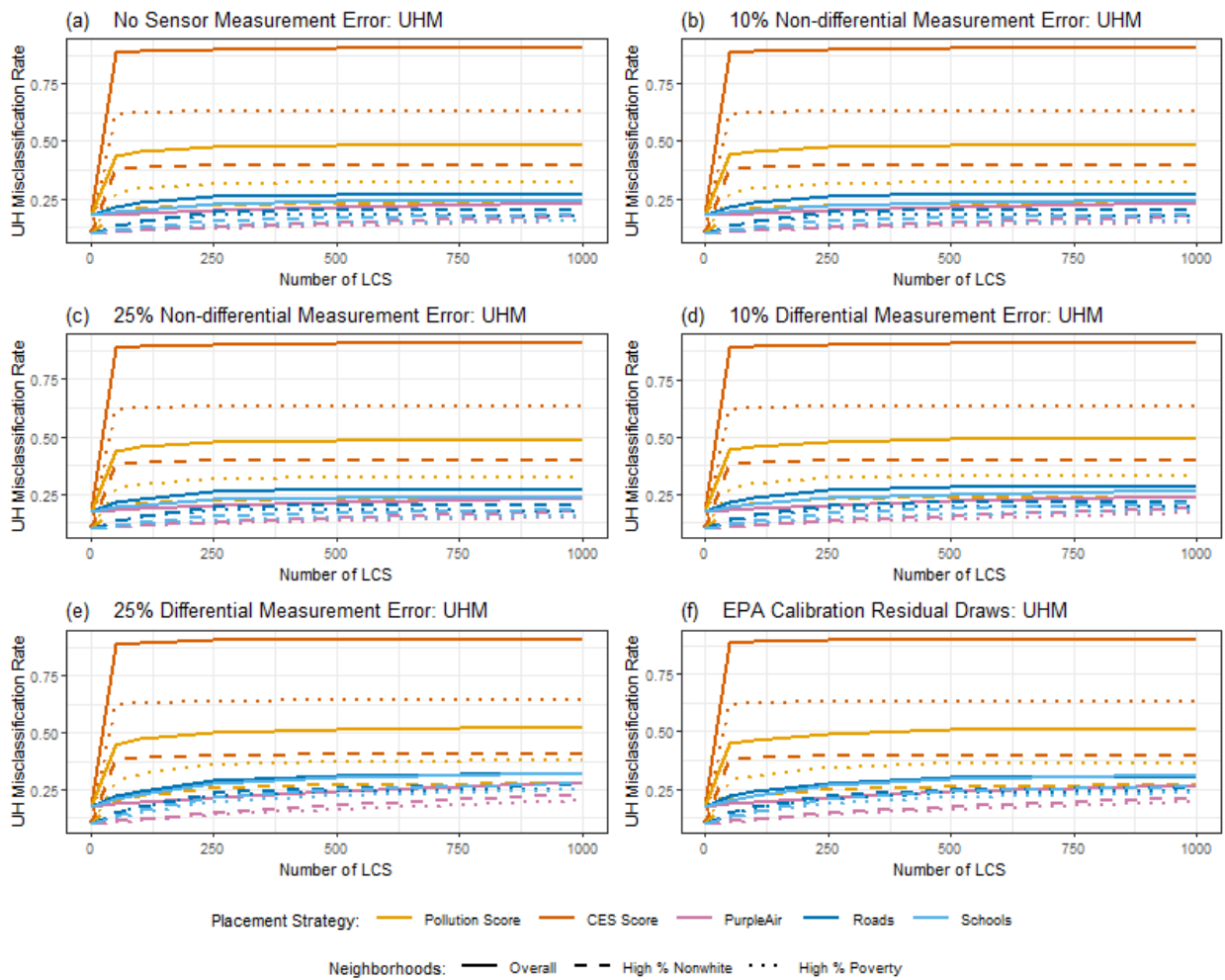

**Fig. S9. Unhealthy-Healthy Misclassifications, unweighted by population density.** Rates of UH misclassification resulting from different numbers of LCS deployed, LCS placement strategies, and sensor measurement error types and amounts. UH misclassification occurs when the air quality is unhealthy (Orange+) but is reported as healthy (Green or Yellow); the UHM rate is calculated by dividing the fraction of UH misclassifications by the total fraction of unhealthy days experienced by each group. All results were calculated using 366 days and averaged across 100 simulation replicates, unweighted by population density.

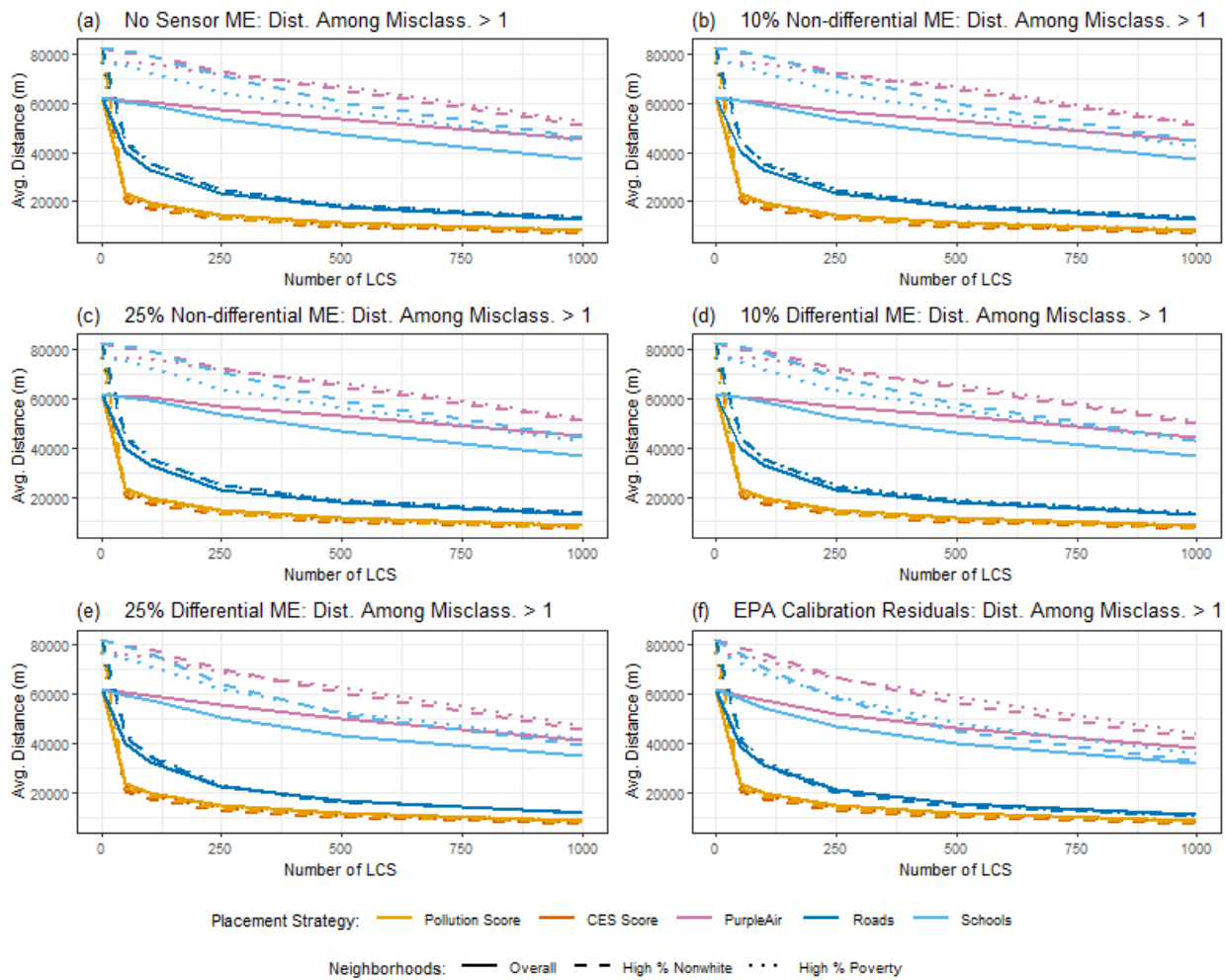

**Fig. S10. Distance to the nearest monitor or sensor, among observations misclassified by more than one level of the AQI** (e.g. Red to Yellow, or Green to Orange). All results were calculated using 366 days and averaged across 100 simulation replicates, **unweighted by population density**.

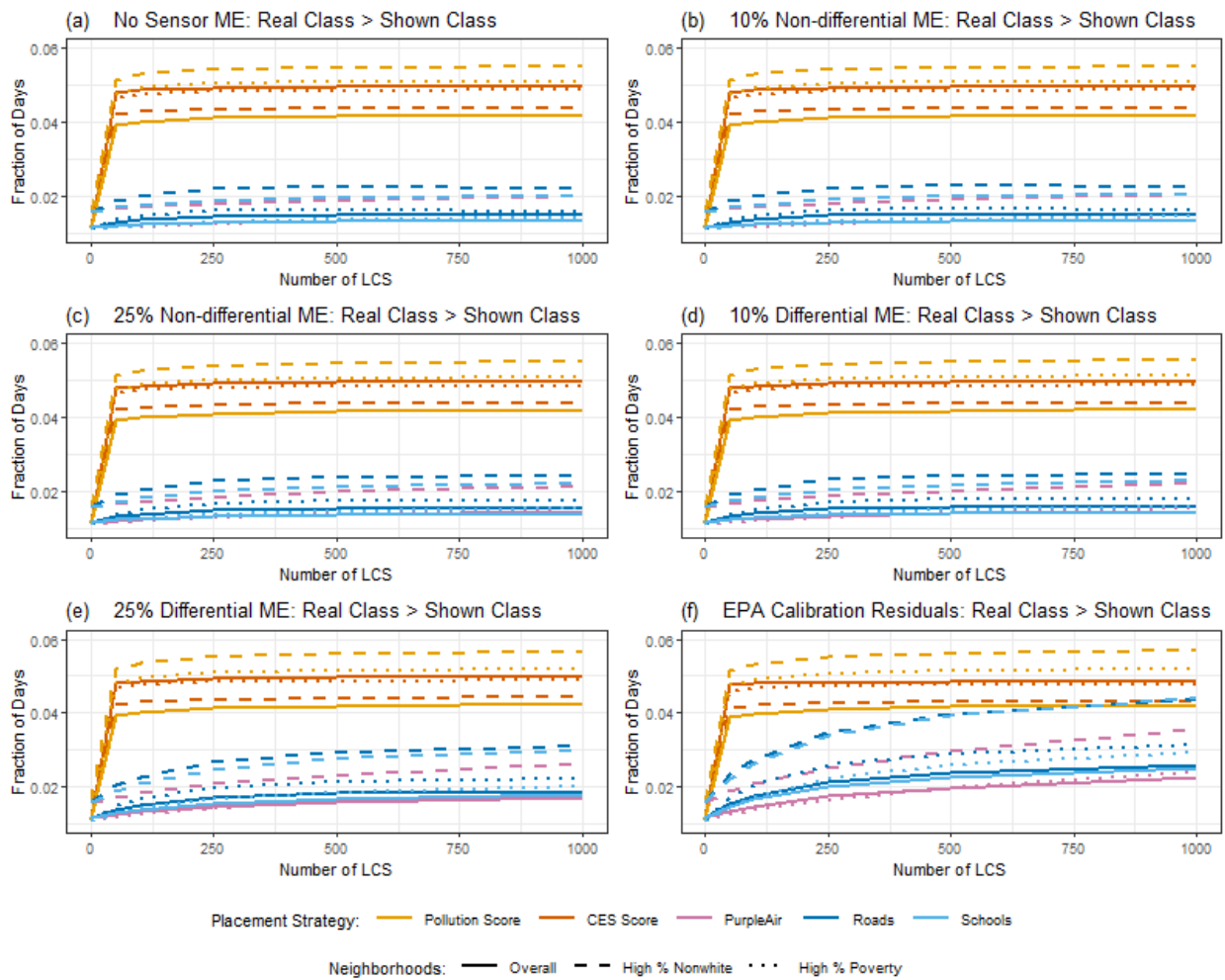

**Fig. S11. Underclassifications of the AQI, unweighted by population density.** The average fraction of days on which the real AQI class (what was truly experienced) was higher than the shown AQI class (reported from the nearest monitor or sensor), resulting from different numbers of LCS deployed, LCS placement strategies, and sensor measurement error types and amounts. All results were calculated using 366 days and averaged across 100 simulation replicates, unweighted by population density.

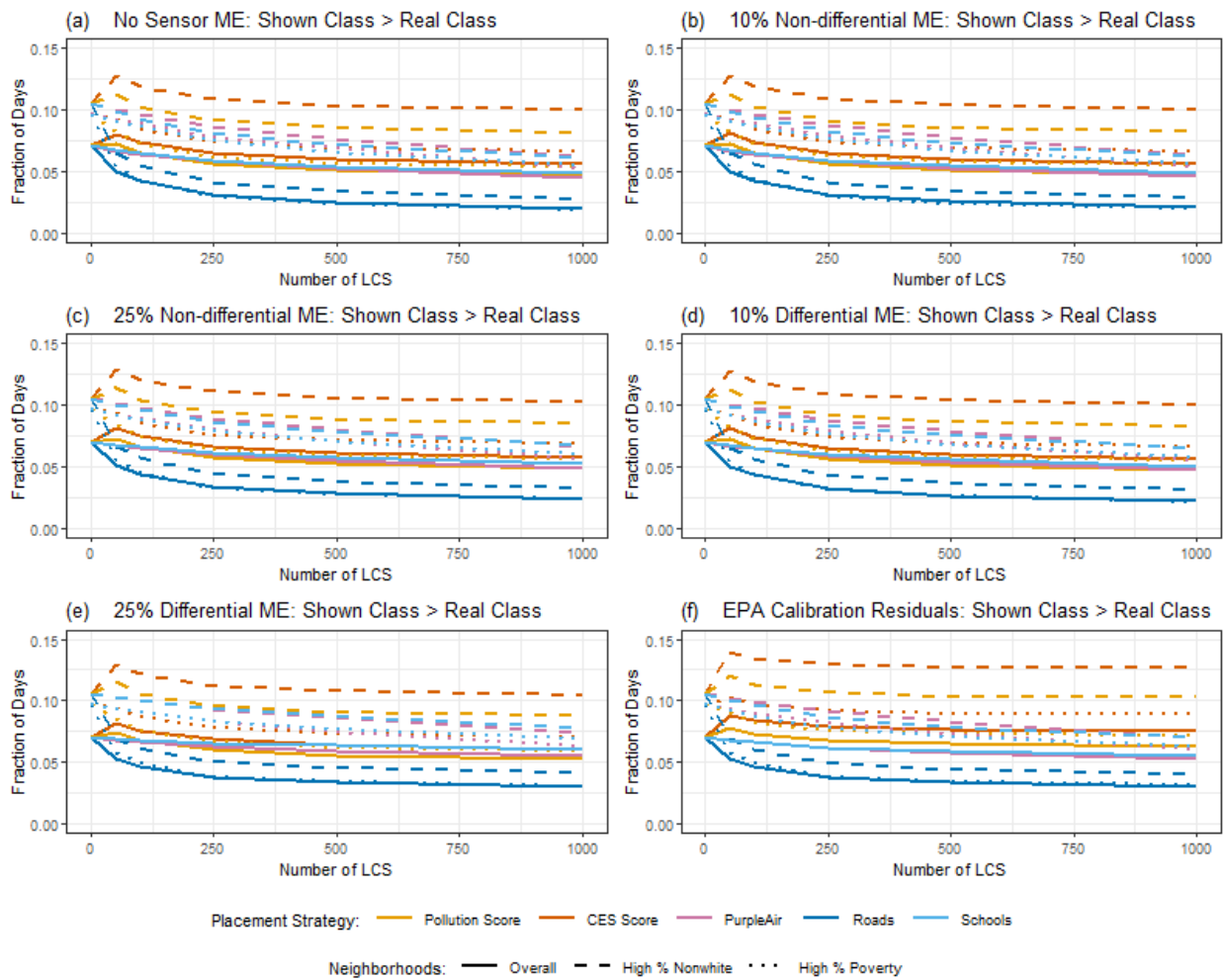

**Fig. S12. Overclassifications of the AQI, unweighted by population density.** The average fraction of days on which the shown AQI class (reported from the nearest monitor or sensor) was higher than the real AQI class (what was truly experienced), resulting from different numbers of LCS deployed, LCS placement strategies, and sensor measurement error types and amounts. All results were calculated using 366 days and averaged across 100 simulation replicates, unweighted by population density.

**Table S9: Comparing impacts of different sensor accuracies for LCS at all schools**, unweighted by population density (counterpart of Table S6). Results when there are no LCS vs. LCS at all school locations (n = 7,548), assuming different kinds and amounts of sensor measurement error. Results are the average of 100 simulation replicates to account for randomness in the sensor measurement error generation. Unless otherwise specified, "errors" refer to the difference between the true exposure experienced at each grid centroid and the exposure reported from the nearest AQ instrument. AQI under-classification is when the true exposure class is greater than what someone is shown, and over-classification is when the true exposure class is less than what someone is shown. Rate of UH misclassification (UHM) is the fraction of days with unhealthy AQI (Orange+) that are misreported as healthy AQI (Green or Yellow).

| Type/Amount of Sensor Measurement Error | Std. Dev. of Sensor Measurement Error ($\mu g/m^3$) | MAE ($\mu g/m^3$) | 95th Percentile of Errors ($\mu g/m^3$) | Under-classified AQI (%) | Over-classified AQI (%) | UHM (%) |
|---|---|---|---|---|---|---|
| Overall Population | | | | | | |
| No LCS (only EPA monitors) | — | 2.51 | 8.08 | 1.17 | 7.09 | 17.67 |
| No Sensor Error | 0 | 1.78 | 5.74 | 1.35 | 3.40 | 21.91 |
| 10% Non-differential | 0.5 | 1.84 | 5.80 | 1.38 | 3.50 | 21.91 |
| 25% Non-differential | 1.25 | 2.08 | 6.10 | 1.48 | 3.96 | 22.16 |
| 10% Differential | 0.93 | 1.86 | 5.87 | 1.53 | 3.70 | 25.05 |
| 25% Differential | 2.33 | 2.14 | 6.63 | 1.91 | 4.83 | 32.76 |
| EPA Correction Residual Decile Draws | 3.54 | 2.71 | 7.34 | 2.78 | 4.55 | 31.36 |
| Population living in CBGs with high % nonwhite | | | | | | |
| No LCS (only EPA monitors) | — | 2.81 | 9.43 | 1.58 | 10.56 | 9.86 |
| No Sensor Error | 0 | 1.72 | 5.60 | 1.95 | 3.65 | 15.16 |
| 10% Non-differential | 0.5 | 1.80 | 5.66 | 2.04 | 3.84 | 15.57 |
| 25% Non-differential | 1.25 | 2.06 | 5.98 | 2.29 | 4.53 | 16.10 |
| 10% Differential | 1.01 | 1.86 | 5.77 | 2.38 | 4.18 | 19.82 |
| 25% Differential | 2.53 | 2.28 | 6.79 | 3.24 | 5.83 | 29.46 |
| EPA Correction Residual Decile Draws | 3.79 | 2.81 | 7.76 | 4.98 | 5.37 | 27.90 |
| Population living in CBGs with high % poverty | | | | | | |
| No LCS (only EPA monitors) | — | 2.95 | 9.40 | 1.11 | 9.63 | 9.89 |
| No Sensor Error | 0 | 1.75 | 5.79 | 1.46 | 3.20 | 12.63 |
| 10% Non-differential | 0.5 | 1.82 | 5.84 | 1.51 | 3.33 | 12.97 |
| 25% Non-differential | 1.25 | 2.07 | 6.13 | 1.65 | 3.87 | 13.42 |
| 10% Differential | 1.03 | 1.84 | 5.93 | 1.71 | 3.57 | 16.97 |
| 25% Differential | 2.58 | 2.17 | 6.75 | 2.25 | 4.91 | 27.11 |
| EPA Correction Residual Decile Draws | 3.86 | 2.76 | 7.50 | 3.39 | 4.63 | 25.37 |